\documentclass{article}
\usepackage[utf8]{inputenc}
\usepackage[colorlinks=true, allcolors=blue]{hyperref}

\usepackage{commands}
\usepackage{computed_values}

\title{A Large Scale Survey of Motivation in Software Development and Analysis of its Validity}

\author{Idan Amit ~~~~~~~~~~~~~~~~~~~~~~ Dror G. Feitelson\\
idan.amit@mail.huji.ac.il ~~~~~~~~~ feit@cs.huji.ac.il\\
Department of Computer Science \\
The Hebrew University, Jerusalem 91904, Israel
}

\date{}

\begin{document}

\maketitle

\begin{abstract}
\noindent\textbf{Context:}
Motivation is known to improve performance.
In software development in particular, there has been considerable interest in the motivation of contributors to open-source.

\noindent\textbf{Objective:}
We identify 11 motivators from the literature (enjoying programming, ownership of code, learning, self-use, etc.), and evaluate their relative effect on motivation.
Since motivation is an internal subjective feeling, we also analyze the validity of the answers.

\noindent\textbf{Method:}
We conducted a survey with 66 questions on motivation which was completed by \completedNum developers.
Most of the questions used an 11-point scale.
We evaluated the answers' validity by comparing related questions, comparing to actual behavior on GitHub, and comparison with the same developer in a follow-up survey.

\noindent\textbf{Results:}
Validity problems include moderate correlations between answers to related questions, as well as self-promotion and mistakes in the answers.
Despite these problems, predictive analysis---investigating how diverse motivators influence the probability of high motivation---provided valuable insights.
The correlations between the different motivators are low, implying their independence.
High values in all 11 motivators predict increased probability of high motivation.
In addition, improvement analysis shows that an increase in most motivators predicts an increase in general motivation.

\noindent\textbf{Conclusions:}
All 11 motivators indeed support motivation, but only moderately.
No single motivator suffices to predict high motivation or motivation improvement, 
and each motivator sheds light on a different aspect of motivation.
Therefore models based on multiple motivators predict \emph{motivation improvement} with up to 94\% accuracy, better than any single motivator.

\end{abstract}

\paragraph{Keywords:}
Motivation,
Software engineering,
Open-source development,
Survey validity.

\hide{
\section{Statements and Declarations}
We declared that we do not have any financial or non-financial interests that are directly or indirectly related to the work.
This research was supported by the ISRAEL SCIENCE FOUNDATION (grant no.\ 832/18).

\section{Author contributions}
Both authors contributed to the study conception, design and analysis. Data collection and coding were performed by Idan Amit. The first draft of the manuscript was written by Idan Amit and both authors wrote and edited previous versions of the manuscript. Both authors read and approved the final manuscript.

\newpage
}
\section{Introduction}\label{sect:introduction}

\doit{open replication package}

Motivation has a high impact on human performance in many fields \cite{Frangos1997, ariely2008man, LAM20081109, Wright00Satisfaction, Judge01Satisfaction}.
In the context of software development it is especially interesting \cite{DBLP:journals/infsof/BeechamBHRS08, Sharp:2009:IIS:1572193.1572221},
due to the phenomenon of open-source development \cite{Raymond:1999:CB:580808}, where many of the developers are volunteers \cite{DBLP:journals/infsof/SharpBBHR09}.

We conducted a large-scale survey in order to investigate how developers see the motivators affecting their motivation.
Our survey contained questions related to eleven motivators which may affect motivation, taken from prior work.
Some of these motivators relate to the culture of open source, such as adhering to its \textbf{ideology} \cite{belenzon2008motivation} and developing software for \textbf{self-use} \cite{Raymond:1999:CB:580808}.
Other motivators are related to internal reasons, such as \textbf{enjoyment} from developing code \cite{Graziotin2017Happy, Argyle1989Happy}, \textbf{learning} from it \cite{Bartol:1982:MIS:2025521.2025525}, the feeling of overcoming a \textbf{challenge} \cite{MCCLELLAND1961Achieving,LOCKE1968157}, or gaining \textbf{ownership} of a project \cite{LAM20081109, BADDOO200285}.
Additional motivators extend this to a wider context: being part of a \textbf{community} \cite{ Asproni2004,Ye:2003:TUM:776816.776867}, receiving \textbf{recognition} for one's work \cite{gerosa2021shifting, VonKrogh:2012:CRM:2481639.2481654} --- or, in unpleasant situations, suffering from \textbf{hostility} \cite{Raman2020StressAB, Adams1963TOWARDAU}.
Finally, there are project-based motivators like a sense of \textbf{importance} \cite{Hackman1976JCT}, as well as receiving \textbf{payment} for participating in a project \cite{benabou2003intrinsic, riehle2007economic}.

The survey contained 66 questions, covering motivation, satisfaction, demographics, and self-rating of skill.
We obtained answers from \devNum developers, and \completedNum of them completed the whole survey.
A year later we conducted a followup survey, answered by \followupParticipants of the original participants.
We used the answers to measure the correlation of each motivator with general motivation, evaluate motivators as predictors of general motivation, evaluate how improvement in a motivator improves general motivation, and build a motivation model to measure the overall predictability.
Taken together, a motivator that is influential in all these separate ways is likely to have a true impact.

In general, all the motivators were found to contribute to motivation in the predictive sense (knowing of a high motivator means higher probability of motivation).
Other than challenge, ideology, and hostility, an improvement in the follow-up answers of the motivator also increases the probability of motivation improvement.
On the other hand, none of the motivators is sufficient or necessary on its own for high motivation.
Yet, when used together in a predictive model, one can predict well both high motivation and motivation improvement.

Motivation is an internal subjective feeling.
When answering regarding our motivation, a valid answer depends on internal identification of the motivation, correct translation into a proper answer, and lack of personal biases \cite{BassettJones2005Herzberg, Kruger99unskilledand}.
We therefore also analyzed the validity and reliability of the answers.
We did this by examining the correlations between questions concerning the same motivator.
We also checked the consistency of answers of the same person to the same questions in the original and follow-up surveys.
Moreover, we investigated the reliability of the answers, using identification of errors, differences between perception and actual behavior on GitHub, and biases.

We found that the attitudes toward motivators are only moderately stable.
The validity also suffers from many kinds of problems.
Simple mistakes, like typos, are rather rare.
Biases are more common.
For example, only 5.6\% of the participants place themselves in a lower-than-neutral level of `My code is of high quality'.
This may indicate that our participants are indeed highly skilled developers.
However, the GitHub profiles provided by some participants allowed us to compare their answers to their actual behavior.
This showed that the participants overestimated their code quality, their documentation level, and their productivity.
Despite these problems, questions belonging to the same motivator are reasonably correlated.
Comparing answers of the same developer in the original and follow-up survey also shows a moderate stability.
Therefore, the results that were supported by multiple analyses seem to be valid.

This study makes three main contributions:
\begin{itemize}
\item 
We conducted a large-scale survey on software developers' motivation, covering 11 motivators.

\item
We make several methodological innovations, including about assessing the validity of the results:
\begin{itemize}
\item
We asked participants for their GitHub profiles, which enabled comparing survey answers and actual behavior.
\item
We conducted a follow-up survey, asking the same people the same questions again after more than a year.
This allowed us to measure the answers' stability and the impact of changes in motivators on changes in motivation.
\item
We framed the analysis as a supervised learning problem.
We initially considered each single motivator as a classifier for high motivation, moved to full models, and then applied the same methods for motivation improvement.
\item 
We used the correlations of different questions in the same topic to measure absolute and relative coherence.
\item
The large scale of the study allowed us to compare answers of different people in the same project, estimating subjectivity.
\item
We used several types of analyses in tandem to investigate the relations between motivators and motivation.
Most relations were consistent in most or all the methods, testifying to their validity.
\end{itemize}

\item
We analyze the results using a machine-learning framework and reach several conclusions regarding motivators and their influence:
\begin{itemize}
\item We corroborate previous work showing that in general motivators from prior work are indeed correlated with motivation.
\item At the same time we find that none of them alone is enough to guarantee motivation, so developers usually need several reasons in order to have high motivation.
Also, predictive performance is improved by taking multiple motivators into account.
\item Answers regarding motivation have moderate validity, shown by comparing to similar questions, comparing to a different date, or comparing to actual behavior.
\item We found that although hostility is rare, when it exists it has a negative influence on motivation.
Yet, it tends to be unobserved by others in the same project.
\end{itemize}
\end{itemize}

\section{Related Work}

\idan{add theories and general work}

\subsection{Motivation}

Motivation, in general and in work context, has been extensively investigated due to its importance.
Many theories were suggested.
Skinner suggested the operant conditioning, learning behavior due to reward and punishments \cite{Skinner1938OperantConditioning}.
Maslow's hierarchy of needs sees self-actualization as the top need \cite{Maslow1943Motivation}.
McClelland argues that motivation comes from a mixture of affiliation (society based), authority (opportunities to gain it), and achievements (overcoming challenges) \cite{MCCLELLAND1961Achieving}.
The equity theory claims that motivation might be hurt due to relative comparison and the feeling of not being fairly treated \cite{Adams1963TOWARDAU}.

In the context of work, the Goal Setting Theory claims that challenging yet achievable goals benefit the motivation \cite{LOCKE1968157}.
Close in spirit is Vroom's Expectancy Theory \cite{Vroom1964Expectancy} that claims that one estimates the outcome, the outcome value, and the probability of the value.
Given these, the motivation is determined, and one will have more motivation in tasks where an outcome that is valued is likely to be achieved.

Herzbereg et al.\ suggested the Motivation-Hygiene Theory \cite{Herzberg1959Motivation}.
According to it, positive motivation is usually due to intrinsic motivators.
However, external hygiene factors might lead to the loss of motivation.
Hackman and Oldman suggested the Job Characteristics Theory \cite{Hackman1976JCT}.
They claim that the motivation might come from the job itself, due to the significance, autonomy, skill, identity, and feedback related to the job.

These theories are classical and were introduced long ago.
Though they were criticized, they are still beneficial \cite{BassettJones2005Herzberg}.

The relation between job satisfaction and performance is an important research area in organizational psychology \cite{Roethlisberger39Hawthorne, Landy89Work, Judge01Satisfaction}.
Plenty of empirical work was done in order to understand motivation.
Task significance, a motivation cause, was shown to increase productivity \cite{Grant2008TaskSignificance}.
Campbell et al.\ see performance as a function of motivation but also of knowledge and skill \cite{campbell93}.

Job satisfaction also correlated with performance in field studies \cite{Wright00Satisfaction}.
Judge et al.\ performed a meta-analysis and reached 0.3 correlation from 312 studies with 54,417 participants \cite{Judge01Satisfaction}.

\subsection{Motivators}\label{sect:related-motivation-factors}

We investigate motivators in software development and specifically in open-source development.
Some of these motivators are relevant to any human activity, others to software development or open-source development.
Demarco and Lister \cite{demarco:people}, and also Frangos \cite{Frangos1997}, claim that the important software problems are human and not technological.
So, many have investigated motivation in software engineering \cite{LENBERG201515,DBLP:journals/infsof/BeechamBHRS08,  8370133, Ghayyur2018}.

Open-source development is the collaborative development of software that is free to use and further modify.
The best-known non-technical equivalent is Wikipedia.
A seminal description of the phenomenon is given in Raymond's ``The Cathedral and the Bazaar'' \cite{Raymond:1999:CB:580808}.
Payment is probably the most common way to motivate people to perform a task, though it is an extrinsic motivation and therefore its influence is more complex \cite{benabou2003intrinsic, riehle2007economic}.
However, it is common to perform open source software development as a volunteer, which means that salary is not the motivation, making it startling from an economical point of view at first sight \cite{Josh2002Economics}.
Therefore, the motivation of open source developers was investigated as a specific domain, in an effort to uncover other motivators \cite{Ye:2003:TUM:776816.776867, VonKrogh:2012:CRM:2481639.2481654, Cristina04altruisticindividuals, Li2006Motivating, HERTEL20031159, Roberts:2006:UMP:1246148.1246151}.
In Section \ref{sect:motivating-factors} we list and discuss the 11 motivators used in our study.

\hide{

In the context of work motivation, the Goal Setting Theory claims that challenging yet achievable goals benefit the motivation \cite{LOCKE1968157}. 
Indeed, research has shown that developers are motivated by technical challenges and overcoming them \cite{Sharp:2009:IIS:1572193.1572221, 1252263, Ramachandran:2006:ETI:1125170.1125221}.

Hostility is a general demotivator \cite{Raman2020StressAB, Adams1963TOWARDAU}, aligned with the  Motivation-Hygiene Theory of Herzbereg et al. \cite{Herzberg1959Motivation}.

Two important motivators that have been identified are ownership and autonomy \cite{LAM20081109, BADDOO200285, 8666786, Hackman1976JCT}, especially in building products.

Learning is supported by Bartol and Martin's investigation of information system personnel management \cite{Bartol:1982:MIS:2025521.2025525}.
Recognition \cite{gerosa2021shifting, riehle2007economic, Raymond:1999:CB:580808, VonKrogh:2012:CRM:2481639.2481654},
enjoyment \cite{Graziotin2017Happy, Argyle1989Happy},
importance \cite{Hackman1976JCT},
and community \cite{MCCLELLAND1961Achieving, beggan1992social, Asproni2004, Ye:2003:TUM:776816.776867, 10.1145/3422392.3422433}
are general motivators.
Self-use \cite{Raymond:1999:CB:580808} and  ideology \cite{belenzon2008motivation} are common motivators in open-source development.
}

\subsection{Reliability of Motivation Reports}\label{sect:related-work-methodology}

The limited reliability of motivation reports, a problem that we also cope with, was investigated in prior work.
Using self-estimation in a survey might be a threat to the validity of the collected data.
There might be biases due to ego defenses\cite{BassettJones2005Herzberg}, the Dunning–Kruger effect \cite{Kruger99unskilledand}, subjectivity, and different personal scales. 
Further, ``research on self-esteem (Shavit \& Shouval, 1980) \cite{SHAVIT1980417} has demonstrated empirically that individuals resist lowering favorable self-perceptions'' \cite{10.2307/258579}.
Previous work has tried to evaluate these difficulties.

Argyle \cite{Argyle1989Happy} checked the reliability of self-estimation of happiness and showed it is related to peer and supervisory estimation.
The Maslach Burnout Inventory validated self-estimation on burnout by comparison with the answers of a close person such as a spouse or a co-worker \cite{Maslach1997Burnout}.
Judge et al.\ \cite{Judge1998Dispositional} also compared a person's and significant other's answers.
For work answers ``The average correlation between the self and significant-other reports, corrected for unreliability, was r = .68.''

Wigert and Harter investigated performance reviews, an area close to motivation \cite{Wigert2019Re-Engineering}.
They mention methodological difficulties when one tries to rely on supervisory estimation instead of self-estimation: individual supervisory ratings are a much less reliable measure of performance than objective measures \cite{Viswesvaran1996Comparative}, and 62\% of the variance in ratings can be attributed to rater bias, while actual performance accounts for just 21\% of the variance \cite{Scullen2011Performance}.
Yet, Tsui reports that an employee and his manager's evaluation of effectiveness match \cite{Tsui1982Reputation}.

Beatty et al.\ \cite{doi:10.1111/j.1744-6570.1977.tb02333.x} also compare manager and employee's appraisals.
They found that there is agreement on medium performance and some disagreements on high and low performance.
In a second usage there was higher agreement, though it was not clear if it was due to clarification of requirements or just better communication.

As prior work shows, there is a moderate agreement between self-reports and a close person's report.
This moderate agreement supports the self-reported answers validity yet warns that they are not perfectly accurate.
In this study, we compare the same person's answers to related questions, and the same person's answers in the original and follow-up surveys.
Our results are aligned with the prior work, also indicating moderate correlation.
We also note that despite all the above concerns, Scott et al.\ report that Facebook found that surveys are twice more accurate than predictive analytics in employee churn \cite{Scott2018Surveys}.

\section{The Survey Instrument}

\subsection{Design}

Our goal is to investigate motivators and motivation.
But motivation is not a well-defined concept.
To increase validity, we used questions endorsed by prior work and questions that can be compared to actual activity.
Some of the motivators were represented by multiple questions, to enable internal coherence validation.
This led to the construction of a relatively long survey with 66 questions (see Appendix \ref{sect:survey-questions}).
The first section of the survey was ``Questions regarding yourself'' (18 questions).
This part included general questions about motivation and verifiable questions about conduct (e.g., the writing of detailed commit messages).
We also asked questions about self-rating of skill.

The second section was ``Questions regarding activity in a repository'' (28 questions), in which we asked about a specific project and its related behavior.
This included our motivation ground truth question: ``I regularly have a high level of motivation to contribute to the repository'' (based on \cite{47853}).
It also included questions on eleven motivators from prior work: community, ownership, self-use, etc.

This was followed by the Job Satisfaction Scale questionnaire (10 questions) \cite{Hills2011Validation} as is.
Job satisfaction is assumed to be related to motivation and its motivators.

In the discussion below we note prior work on each motivator (Section \ref{sect:motivating-factors}), and in the replication package we identify the specific source of each survey question \cite{replication}.
By using questions from prior work, we benefit from a previous validation.
Murphy-Hill et al.\ investigated developer productivity \cite{47853}.
We took a productivity question from there and also based a motivation question on the same pattern.
Amabile et al.\ developed a survey for intrinsic and extrinsic motivation, from which we took some questions \cite{Amabile94thework}.
We took questions regarding hostility from \cite{Cortina2001Incivility} and \cite{Kluger17FLS}.
Kuusinen at el.\ investigated flow, intrinsic motivation, and developer
experience, from which we took questions on the topic \cite{10.1007/978-3-319-33515-5_9}.

We also added a ``Demography'' section (8 questions).
Demography is of interest on its own and enables us to compare to prior work.
Last, we ask an open question requesting comments Whose goal is to ensure that we did not miss a significant factor in the structured questions.

An important goal of the survey was to enable us to compare answers regarding motivation and actual behavior.
For example, we could compare answers to ``I write detailed commit messages'' and actual commit message length.
Therefore, we asked the participants to choose a specific project and provide its name, preferably a public GitHub project.
We asked for their GitHub profile for investigating the developer behavior.
We also asked for the email from participants who were interested in the research results (and offered them a gift lottery).
Email and GitHub profile are personally identifying information, which is usually not collected.
We needed them to match the answers to other data related to the same person, but do not include them with the other experimental materials.
This was approved by our IRB (study 09032020).

The survey was designed to take about 10 minutes.
Most questions used a Likert scale \cite{likert1932technique, joshi2015likert} ranging from 1 to 11.
All participants saw the sections in the same order, yet the questions order within the sections was randomized.
We conducted a pilot to verify the time and make sure that the questions are clear.

\subsection{Execution}

The survey was conducted using the Qualtrics platform from December 2019 to March 2021, including initial pilots.
We obtained \devNum participants, \completedNum of them completed the survey.


GitHub is a platform for source control and code development used by millions of users \cite{GithubUsers}.
We initially focused on 1,530 active public GitHub projects with 500+ commits during 2018, described in \cite{Amit2019Refactoring, Amit2021CCP, amit2021end}.
About 40,000 developers contributed to these projects that year, of which 9,000 contributed more than 12 commits.
We extracted developers' email addresses using the GitHub public email API, fetching the emails of the developers that chose to share them publicly.
We sent emails to 3,255 developers with a public email that had enough commits.
Due to a mistake, we sent multiple emails to developers that worked on multiple suitable projects, and we apologize for that.
We also had a gift card lottery, offering \$50 to three of the participants.
This channel led to 339 participants, which is 20\% of the total.

We also recruited participants in social networks by convenience sampling \cite{etikan2016comparison, acharya2013sampling}, which led to the remaining 80\% of the participants.
We used \href{https://www.reddit.com/}{Reddit}, an online discussions site, as an important source of participants.
Reddit has numerous subreddits, channels dedicated to discussions on specific topics.
It has many channels relevant to programmers such as programming language based, operating system based, tools, etc.
We posted slowly in different subreddits, to get familiar with the community, as posting at a high rate might be considered as spamming.
Each subreddit has different formal and informal rules that should be respected. 
We found the people showed interest in the survey, which led to discussions and upvotes.
These in turn led to more attention to the survey.

As noted above, we asked participants for the name of their project and their GitHub profile.
They provided the names of 484 projects and 303 personal GitHub profiles.
But after posting the survey in social media, we noted that many participants stop at the ``Questions regarding activity in a \textbf{repository}'' section since they do not contribute to a GitHub repository.
This was also accompanied by direct feedback saying that.
Since we were interested in answers about motivation in general and needed the GitHub profile only to link to actual behavior, we changed the questions about ``GitHub repository'' to ``any project'', avoiding this drop.

The original survey ended in March 2021.
A year after the last response, in April 2022, we sent a follow up survey to the \devEmailNum participants that provided their emails in the original survey.
We sent them the name of the project on which they initially answered, and asked to answer on the same project in case that they are still active there and on a different project otherwise.
The questionnaire was the same as the first one, with the additional validation question ``Is it the same project on which you answered last time?''.
In the follow up survey, 124 out of the \devEmailNum participants we reached out to answered (36.3\%).

\hide{Our emails to GitHub developers had 4.4\% response rate, close to the 5\% reported by \cite{huangleaving}.
However, due to a mistake we sent email per project to developers active in more than one project.
This was both annoying and mis-estimated the number of developers.
After fixing the mistake the response rate was 8.1\%, close to
the 9\% reported by Feitelson which used a similar dataset \cite{feitelson2021we}.
We believe that the response rate benefited from the use of the GitHub public email API.
Email is used by GitHub in the development, and therefore they are usually updated (our bounce rate was 1\%).
The API returns the emails of developers that choose to share them publicly, hence more agreeing to cold communication.
We believe that many people found the topic, motivation in software development, exciting.
We anecdotally noticed it in up-votes and comments in social media, responses to the open questions in the survey, and the 19.8\% of the participants that left email wanting to learn the research results.
}

\section{Results}


We work in the framework of supervised learning, trying to predict a concept using a classifier.
The concept, which we try to predict, is high motivation.
The classifiers, which provide us with predictions, are the motivators (e.g., high ownership).
We evaluate how well motivators predict high motivation, using metrics that compare the prediction to the actual motivation.
Interesting metrics are the fraction of those with, say, high ownership who indeed are highly motivated (precision),
the improvement over just the prevalence in the population (precision lift),
and what fraction of the highly motivated who have high ownership (recall).

The analysis of each individual motivator with respect to general motivation provides simple basic results, yet ignores more complex relations like confounders.
We therefore also performed additional, more complex analyses.
We analyzed the relations between motivators to see that this risk is small.
We used the follow-up survey to analyze motivation improvement of each motivator alone.
Last, we built combined models utilising all motivators to avoid risk of confounding and leverage the power of all motivators.

\subsection{General Motivation}\label{sect:general-motivation}

\begin{figure}[!ht]
\centering
\includegraphics[width=0.88\textwidth,trim={0mm 0mm 0 0mm},scale=0.8]{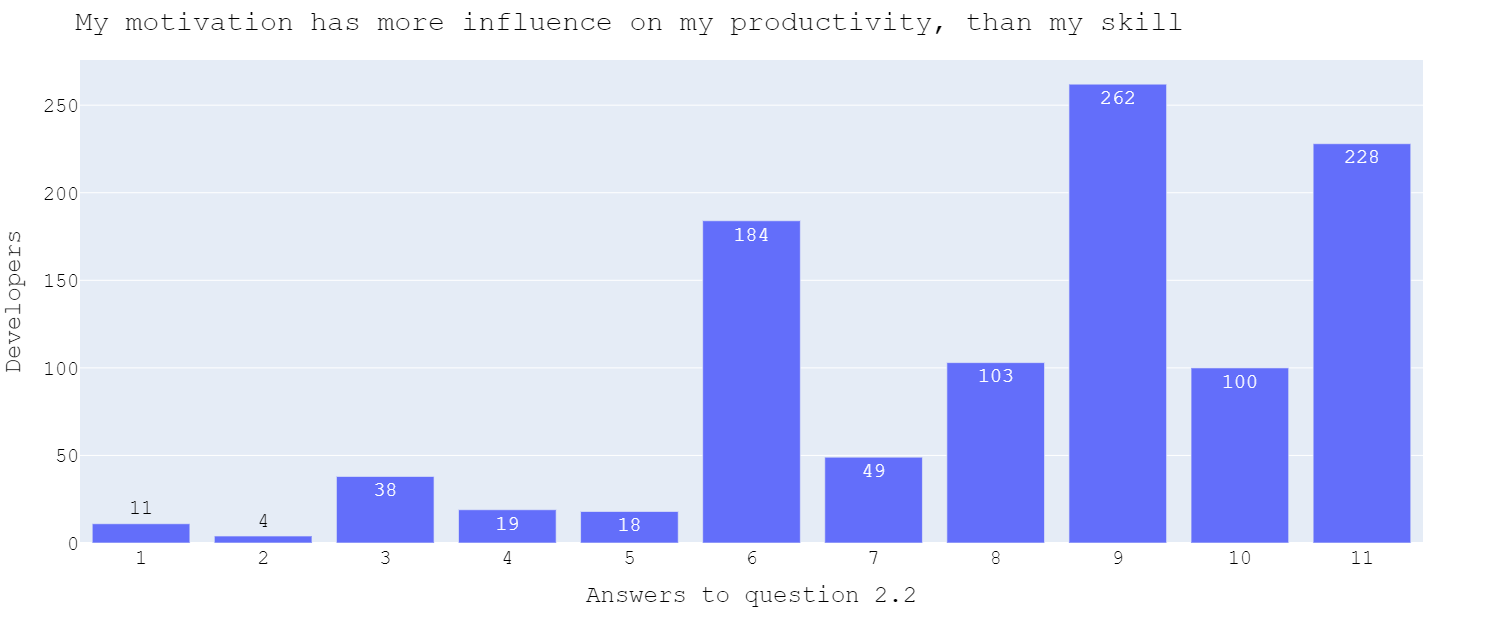}
\caption{\label{fig:2_2_motivation_vs_skill}
Distribution of answers to motivation vs.\ skill question.
}
\end{figure}

Productivity is influenced by many factors, including motivation \cite{shepperd1993productivity}.
In our survey, we explicitly asked about the relative influence of motivation and skill.
73\% of the participants answered that motivation has more influence on their productivity  (answers higher than neutral, see Figure \ref{fig:2_2_motivation_vs_skill}), and only 9\%  answered that their skill is more influential.

Motivation might derive from many motivators, from payment to enjoyment.
The concept that we would like to investigate is high motivation to contribute to a project, and its relations to these various factors.

\begin{figure}[!ht]
\centering
\includegraphics[width=0.88\textwidth,trim={0mm 0mm 0 0mm},scale=0.8]{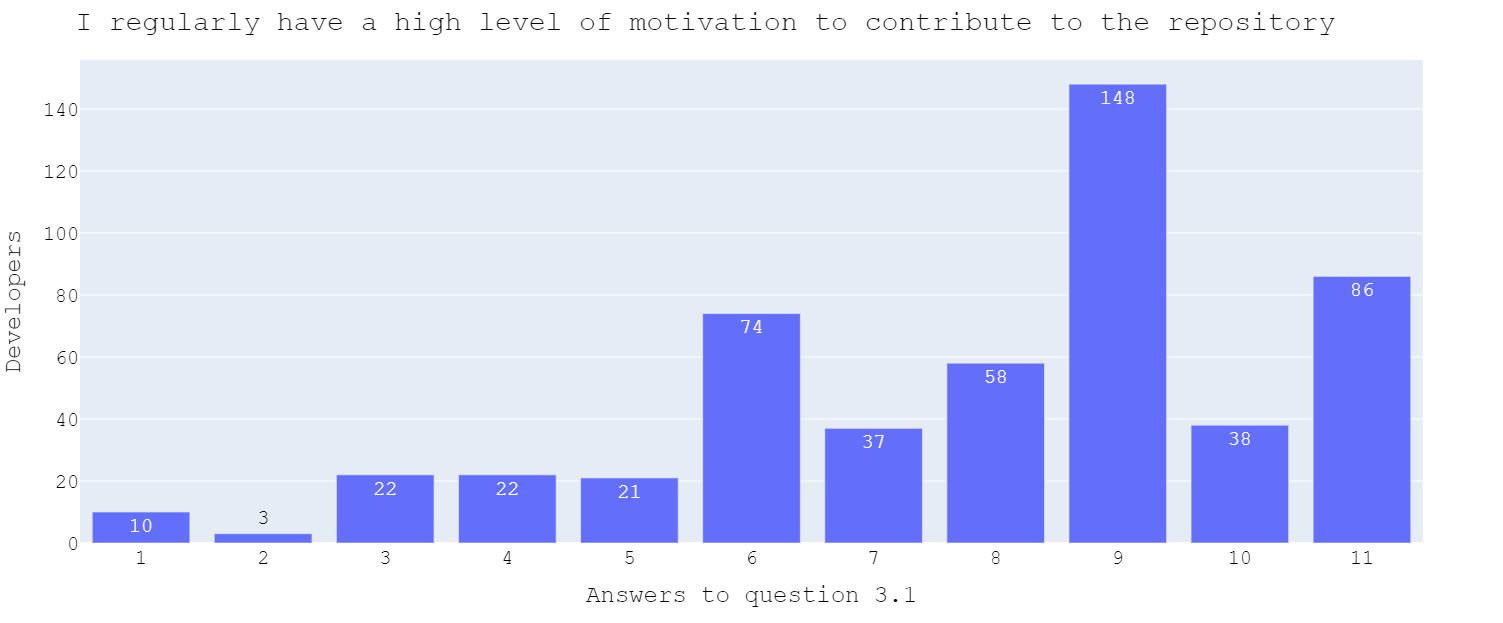}
\caption{\label{fig:high_motivation}
Distribution of answers to general motivation question.
}
\end{figure}

We measure general motivation using the question ``I \textbf{regularly} have a \textbf{high} level of motivation to contribute \textbf{to the repository}'' (pattern is based on \cite{47853}).
The results show that developers are generally motivated (Figure \ref{fig:high_motivation}).
High motivation (at least 9 = `somewhat agree' on a scale of 1 to 11) was reported by 52.4\% of the participants.

We identified paid people by their answer to a specific question about it (question 3.c).
41\% of the participants that answered this question said that they are paid (284 participants).
24\% of the participants (377) that had GitHub in their project name were identified as contributing to it (the question encouraged providing a GitHub project when possible).
Note that being paid and using GitHub are not mutually exclusive.
38\% of participants of GitHub projects that answered the payment question, said that they are paid.

It is generally accepted that motivated workers work longer hours \cite{beckers2004working}.
Showing that our measurement of general motivation exhibits the same relation provides supporting evidence to its validity.
And indeed, participants reporting high motivation (at least 9) reported an average of working 19.3 hours a week on the project, compared to 3.4 hours for those reporting low motivation (below 9).
This result may be tainted by mixing data about paid developers with data about volunteers, who are common in open-source projects.
We checked this by looking at paid workers separately, out of the participants that reported payment, and the influence of payment is indeed large.
For unpaid workers the reported average working hours were 10.8 (high motivation) and 4.5 (low), while for paid workers they were 27.9 (high) and 25.8 (low).
Thus, both paid and unpaid participants work more hours when motivated yet the average of unmotivated paid employees developers is higher than that of motivated unpaid developers and therefore we do not mix them.

We planned to also validate the measurement with the similar question from the Job Satisfaction Survey: ``Taking everything into consideration, how do you feel about your work?'' \cite{Hills2011Validation}.
However, as discussed in Section \ref{sect:answer-validity}, some of the participants were confused and answered most survey questions on an open source project to which they contribute, yet answered the Job Satisfaction Survey on their regular job.
Despite this confusion, the Pearson correlation between the questions is 0.32.
When focusing on paid developers, for which the probability of confusion is lower, the correlation is 0.36.
While this correlation is moderate, it is similar to the inner correlation in the other motivators as shown below.

\subsection{Motivators}\label{sect:motivating-factors}

The research literature has not produced a canonical agreed list of factors that influence motivation.
Mayer et al.\ reviewed 75 years of motivation measures \cite{mayer2007seventy}.
This showed that many different factors have an effect, but the agreement between them is limited.
We therefore needed to select which ones to include in our study.

We based our list of motivators mainly on Beecham et al.'s review of motivation in software engineering \cite{DBLP:journals/infsof/BeechamBHRS08} and Gerosa et al.'s \cite{gerosa2021shifting} work on motivation in open source development.
The motivators that we chose have a long history going back to Herzberg's Motivation-Hygiene Theory \cite{Herzberg1959Motivation}, and therefore were thoroughly investigated over the years (e.g., in general \cite{herzberg1986one} and in software development \cite{couger1988motivators}).
Note that we excluded some of the motivators which are less relevant to open-source development, like ``Job security'' and ``Company policies''.
Conversely, we did include ``hostility'', which is a \emph{de}motivator
(it is common to refer to factors of positive influence as motivators and those of negative influence as demotivators) \cite{Adams1963TOWARDAU}.
Since we have only a single demotivator, we use the term ``motivator'' to refer to both it and the positive motivators.

\hide{
The concept we want to predict is general motivation.
This was operationalized by the answer to the question ``I regularly have a high level of motivation to contribute to the repository'' being 9 `somewhat agree' or above (Section \ref{sect:general-motivation}).
We perform the same categorization for the different motivators, splitting them between answers of less than 9 and answers of 9 and above.
We then use a supervised learning on binary values framework to see whether high motivators predict high motivation.
}

Table \ref{tab:high-answer-motivation-factor} summarizes the predictive performance of all the motivators.
We discuss each of the motivators in the following subsections.
Note that in each row we analyzed participants that answered the motivation question and at least one question about the motivator, so populations are not identical.

We use common metrics used in machine learning and information retrieval.
The concept that we want to predict is high general motivation.
This was operationalized by the answer to the question ``I regularly have a high level of motivation to contribute to the repository'' being 9 `somewhat agree' or above.

Usually in machine learning a classification algorithm (e.g., decision tree) is used to create a model, which is a specific rule providing predictions (e.g., if $A\: \&\:not\:B$).
In contrast, our models are the motivators (e.g., ownership, challenge), also binarized into high and low using 9 as the threshold.
Note that in this part the models are pre-defined, and not learnt by a classification algorithm, and we only evaluate their predictive performance.

The cases in which the concept is true are called `positives' and the positive rate is denoted $P(positive)$ (in our case this is 0.52 as noted above).
Cases in which the model is true are called `hits' and the hit rate is $P(hit)$.
For example, a high hit rate for ownership means that many participants report ownership, and we want to see whether they are also generally motivated.

Ideally, hits correspond to positives, but usually some of them differ.
Precision, defined as $P(positive|hit)$, measures a model's tendency to avoid false positives (FP).
But precision might be high simply since the positive rate is high.
Precision lift, defined as $\frac{precision}{P(positive)} -1= \frac{P(positive|hit) - P(positive)}{P(positive)}$, copes with this difficulty and measures the \emph{additional} probability of having a true positive relative to the base positive rate.
Thus, a useless random model will have precision equal to the positive rate, but a precision lift of 0.
Recall, defined as $P(hit|positive)$, measures how many of the positives are also hits; in our case, this is how many of the highly motivated participants also report high ownership.

\begin {table}[h!]\centering
\caption{ \label{tab:high-answer-motivation-factor} High Motivation Predictability by Motivator }
\begin{tabular}{ | l| c| c| c| c| c | }
\hline
Motivator & Hit rate & \multicolumn{4}{c|}{Performance as predictor of motivation} \\
\cline{3-6}
  & (Fraction $\geq  9$) & Accuracy & Precision & Prec.\ lift & Recall \\
\hline
Enjoyment & 0.74 & 0.64 & 0.62 & 0.18 & 0.86\\ \hline
Ownership & 0.73 & 0.59 & 0.57 & 0.10 & 0.81\\ \hline
Learning & 0.72 & 0.59 & 0.58 & 0.10 & 0.80\\ \hline
Importance & 0.63 & 0.61 & 0.61 & 0.16 & 0.73\\ \hline
Challenge & 0.62 & 0.63 & 0.62 & 0.20 & 0.74\\ \hline
Self-use & 0.56 & 0.60 & 0.61 & 0.17 & 0.65\\ \hline
Ideology & 0.53 & 0.57 & 0.59 & 0.13 & 0.60\\ \hline
Recognition & 0.48 & 0.58 & 0.60 & 0.18 & 0.56\\ \hline
Payment & 0.45 & 0.55 & 0.58 & 0.10 & 0.49\\ \hline
Community & 0.41 & 0.63 & 0.67 & 0.35 & 0.53\\ \hline
Hostility & 0.07 & 0.52 & 0.65 & 0.30 & 0.08\\ \hline
\end{tabular}
\end{table}

We now present our analysis of each individual motivator, from the most to the least prevalent.
We show how common high answers (9 and above) are to each motivator, in general, for paid developers, and for open-source developers.

\subsubsection{Enjoyment}\label{sect:enjoyment}

\begin{figure}[!ht]
\centering
\includegraphics[width=0.88\textwidth,trim={0mm 0mm 0 0mm},scale=0.8]{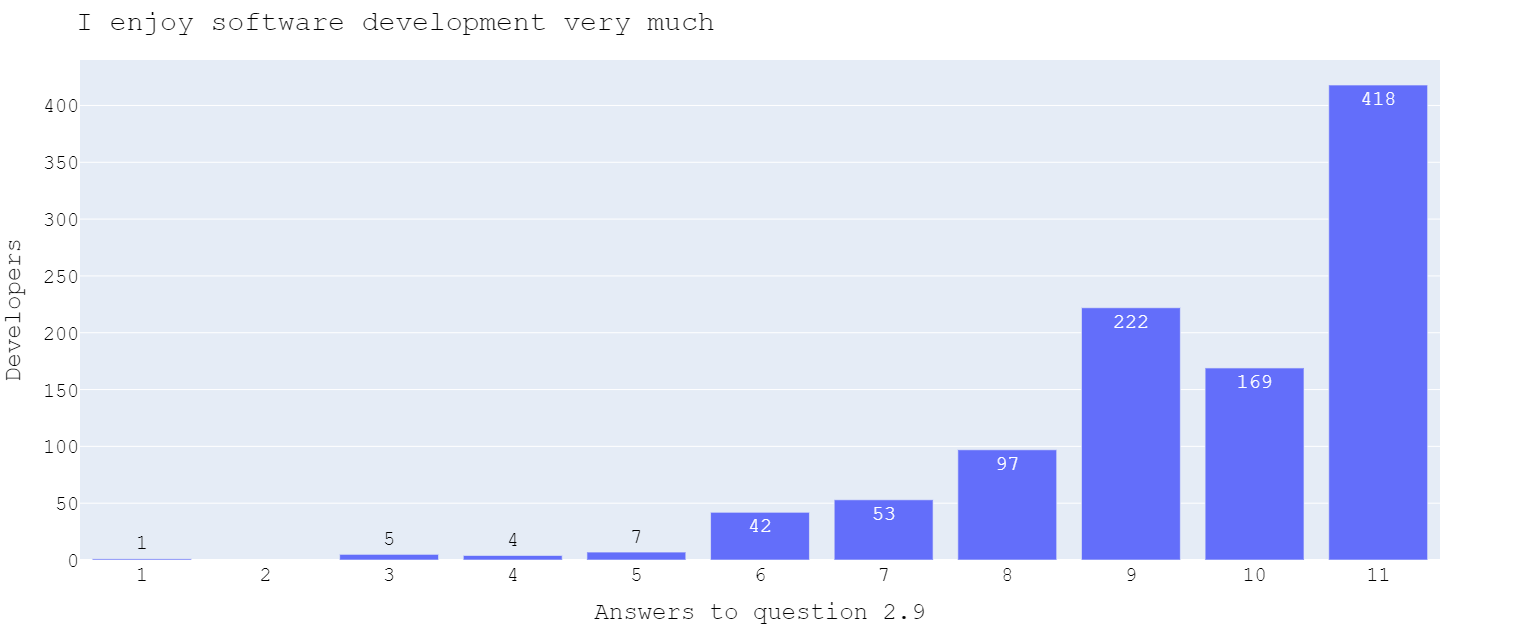}
\caption{\label{fig:enjoy_development}
Answers distribution of enjoyment question.
}
\end{figure}

We measure enjoyment \cite{gerosa2021shifting, Graziotin2017Happy, Argyle1989Happy} by the following questions (numbered by their location in the survey):

\begin{itemize}
\item [2.9] I enjoy software development very much
\item [2.15] I enjoy trying to solve complex problems
\item [3.8] My work on the repository is creative
\item [3.10] I derive satisfaction from working on this repository
\end{itemize}
An example of the results is shown in Figure \ref{fig:enjoy_development}.

We calculated the average answer to all these questions per participant, and then the average of these averages.
This led to an overall average of 9.07.
74\% of the participants reported high enjoyment (at least 9 - `somewhat agree'), more than all other motivators.

76\% of the GitHub participants reported high enjoyment and 75\% of the paid participants.
The correlation of enjoyment with motivation is 0.51, the highest of all motivators.

The recall of enjoyment is 0.83, making it very common among people of high motivation, more than both the hit rate and the positive rate.
The precision is 0.62 and the precision lift is 0.18, which is moderate.

Note that the hit rate of 74\% for enjoyment is significantly higher than the positive rate of 52\%.
This has a large influence on the predictive metrics.
True positives are the intersection of the hits and the positives, so they are bounded by both.
But once the hit rate is higher than the positive rate, the precision is bounded from above.
In our case, the model hit rate is 74\% and the positive rate is 52\%, so the model precision can be at most $\frac{52}{74}=$ 70\%.
The actual precision is 62\% which is not high yet is 89\% of the bound created from the positive rate and the hit rate.

\subsubsection{Ownership}\label{sect:ownership}

We measure ownership \cite{DBLP:journals/infsof/BeechamBHRS08, gerosa2021shifting, LAM20081109, BADDOO200285, 8666786} by the following questions:
\begin{itemize}
\item [3.2] - I have complete autonomy in contributing to the repository
\item [3.3] - I have significant influence on the repository
\item [3.4] - I feel responsible to the repository success
\item [3.16] - I am a core member of the repository
\end{itemize}

The average answer for ownership was 9.02.
73\% of the participants reported high ownership (9 or above, second highest of all motivators).
This was also the percentage for paid participants;
with GitHub participants it was 75\%.
The correlation of ownership with motivation is 0.24.
\hide{
\begin{figure}[!ht]
\centering
\includegraphics[width=0.88\textwidth,trim={0mm 0mm 0 0mm},scale=0.8]{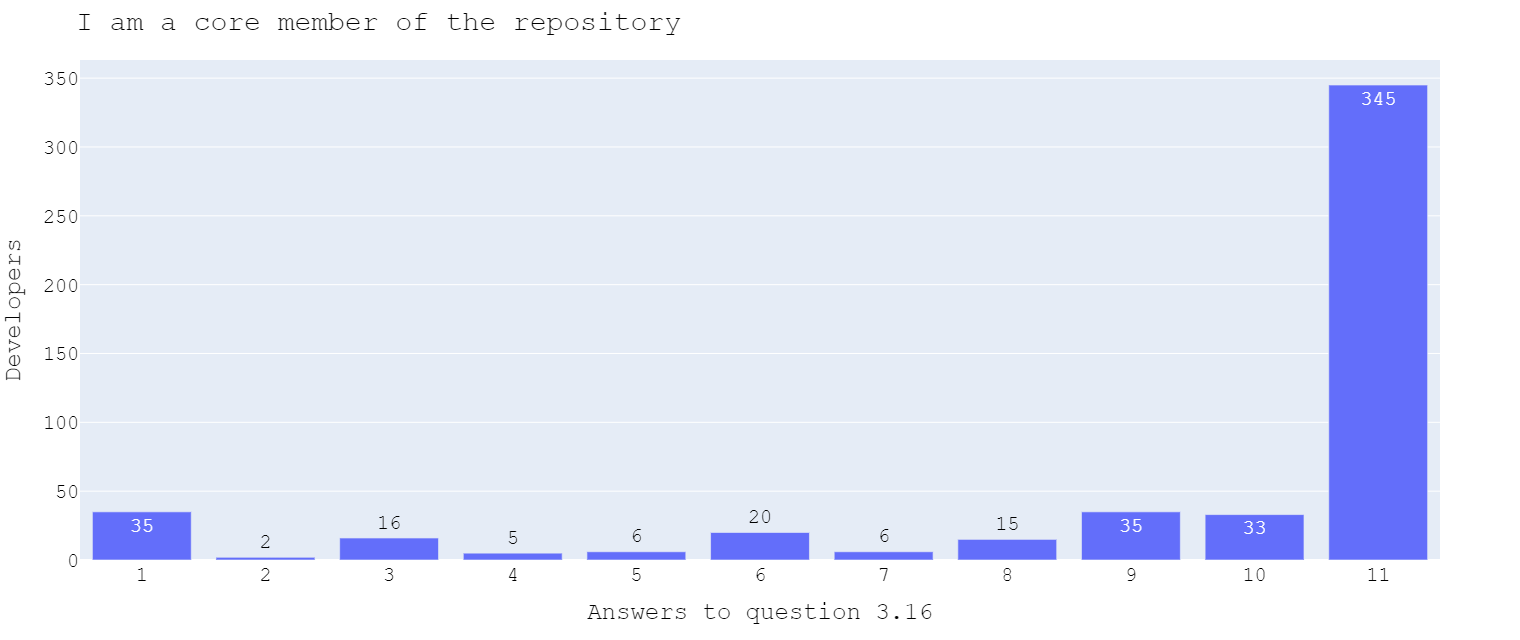}
\caption{\label{fig:3_16_core_member}
Distribution of answers to ownership.
}
\end{figure}
}
The recall when predicting high motivation based on high ownership was 0.81, higher than the hit rate.
However, the precision is 0.57 and the precision lift is only 0.10, partly since ownership is so common.

\subsubsection{Learning}\label{sect:learning}

Learning \cite{DBLP:journals/infsof/BeechamBHRS08, gerosa2021shifting, Bartol:1982:MIS:2025521.2025525} is based on the question:
\begin{itemize}
\item [3.17] - I learn from my contributions
\end{itemize}

The average of the answers was 9.15, the highest among all motivators.
72\% of the participants reported high levels of learning,
70\% of the GitHub participants and 77\% of the paid ones.
The correlation of learning with motivation is 0.23.
\hide{
\begin{figure}[!ht]
\centering
\includegraphics[width=0.88\textwidth,trim={0mm 0mm 0 0mm},scale=0.8]{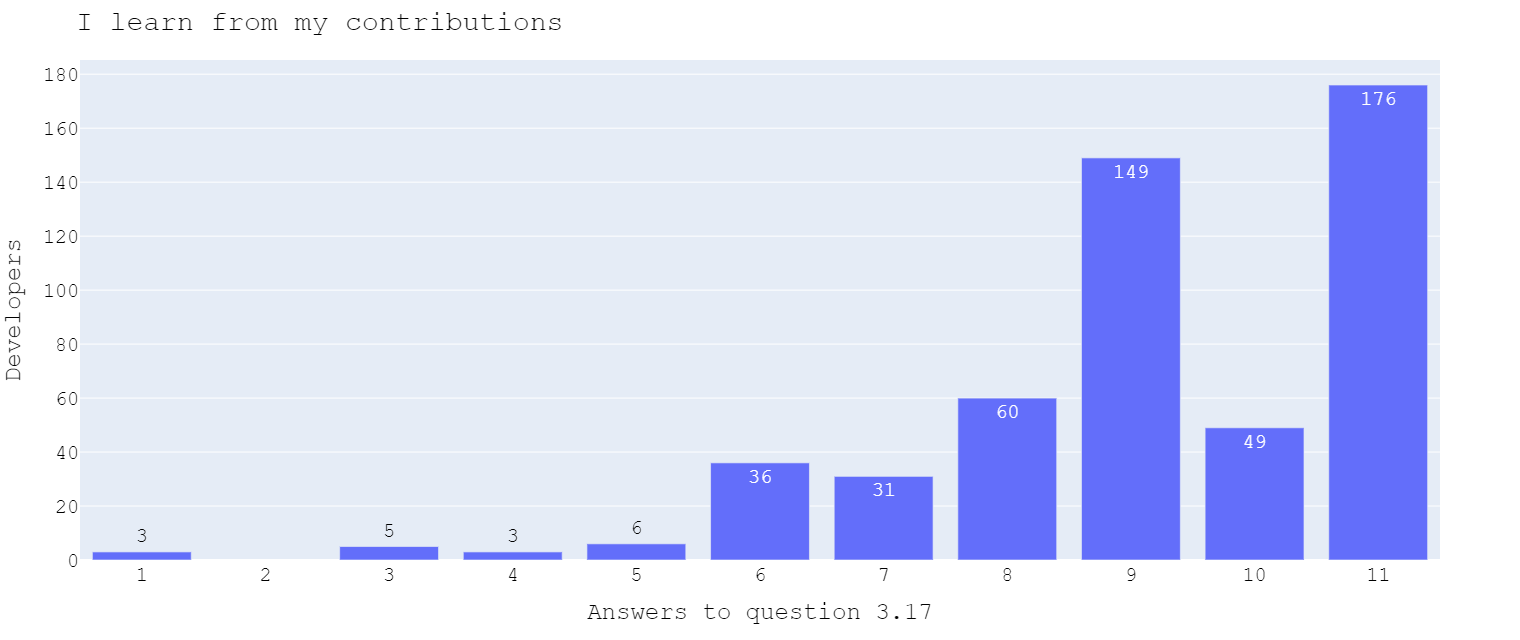}
\caption{\label{fig:3_17_learn}
Answers distribution of learning question.
}
\end{figure}
}
Learning has a recall of 0.80, indicating that it is another very common characteristic of people with high motivation.
Its precision is 0.58 and its precision lift is 0.10, which is relatively low.

\subsubsection{Importance}\label{sect:importance}

Importance \cite{DBLP:journals/infsof/BeechamBHRS08, Hackman1976JCT, Grant2008TaskSignificance} is based on the question:
\begin{itemize}
\item [3.11] - The repository is important
\end{itemize}

The average was 8.62.
The correlation of importance with motivation is 0.35.
63\% of the participants report high feeling of importance.
The same occurred among GitHub participants, compared to 74\% among paid participants.
This is rather surprising since we assume that one will have more considerations and constraints in the context of a paid job than in volunteering to open-source projects.
So, we would assume that one would have higher freedom to choose by importance when volunteering, leading to a higher rate among GitHub participants, but the data shows the opposite.
\hide{
\begin{figure}[!ht]
\centering
\includegraphics[width=0.88\textwidth,trim={0mm 0mm 0 0mm},scale=0.8]{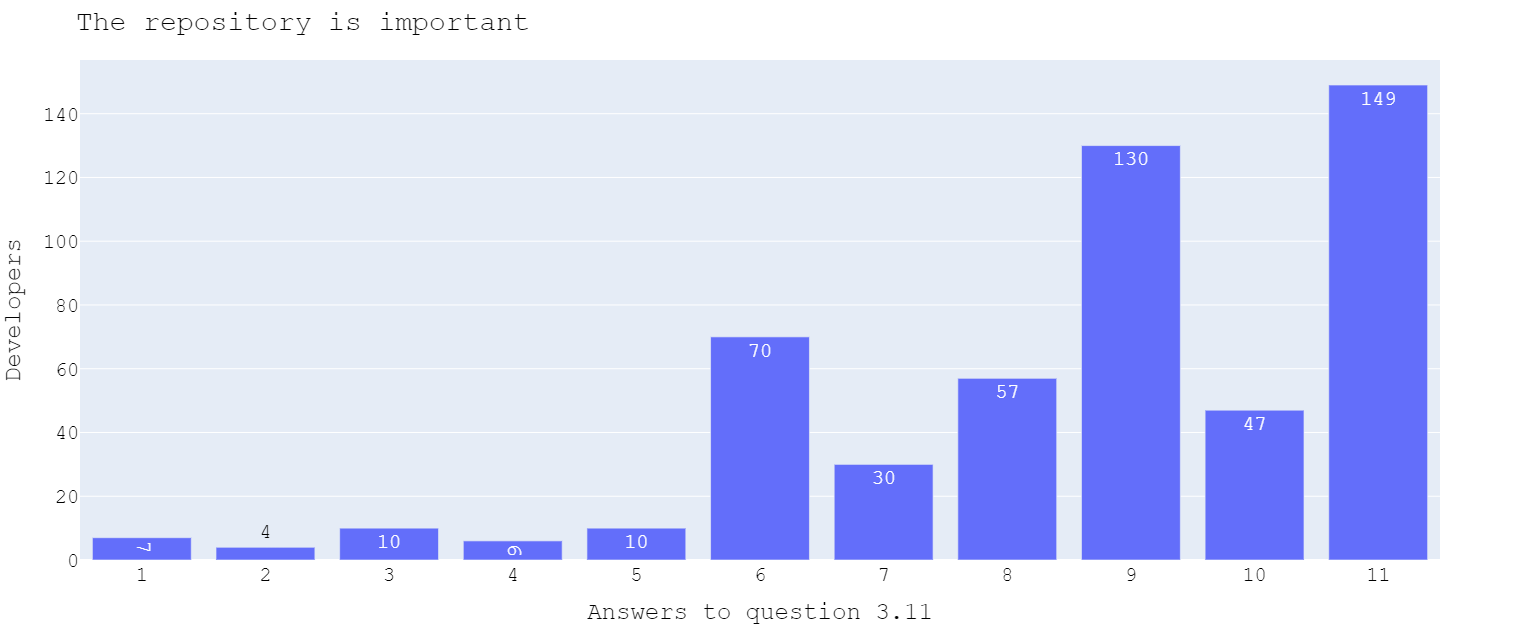}
\caption{\label{fig:3_11_importance}
Answers distribution of importance question.
}
\end{figure}
}
Importance has precision of 0.61 and precision lift of 0.16. 
It has a recall of 0.73, which is high in absolute terms and relative to its hit rate.

\subsubsection{Challenge}\label{sect:challenge}

Challenge \cite{DBLP:journals/infsof/BeechamBHRS08, MCCLELLAND1961Achieving,LOCKE1968157,1252263, Ramachandran:2006:ETI:1125170.1125221,Sharp:2009:IIS:1572193.1572221} is based on the question: 
\begin{itemize}
\item [3.9] - Working on this repository is challenging
\end{itemize}

The average of challenge answers was 8.41. 
62\% of participants reported a high sense of challenge,
60\% of the GitHub participants and 66\% of paid ones.
The correlation of challenge with motivation is 0.30.
\hide{
\begin{figure}[!ht]
\centering
\includegraphics[width=0.88\textwidth,trim={0mm 0mm 0 0mm},scale=0.8]{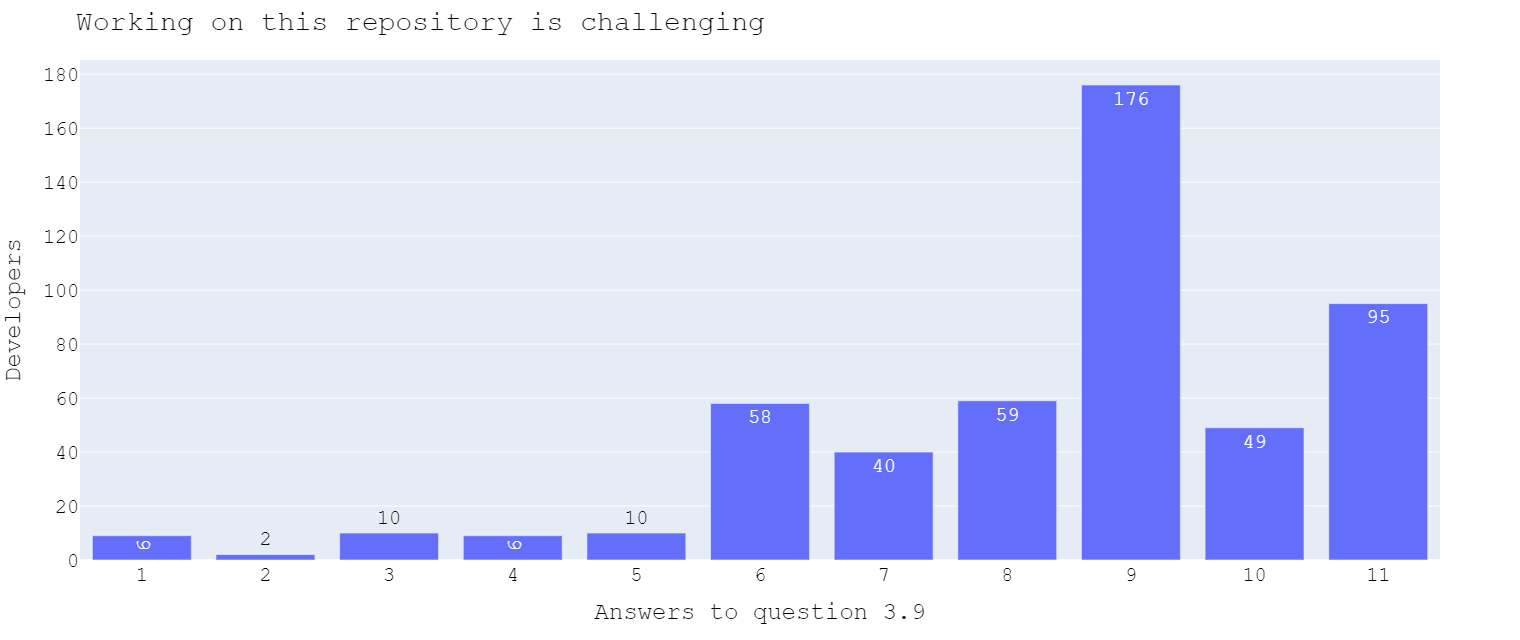}
\caption{\label{fig:3_9_challenge}
Answers distribution of challenge question.
}
\end{figure}
}
Challenge has precision of 0.62 and precision lift of 0.20.
Its recall is 0.74, which is high in absolute terms and relative to its hit rate.

\subsubsection{Self-use}\label{sect:self-use}

Self-use \cite{gerosa2021shifting, Raymond:1999:CB:580808} is based on the question: 
\begin{itemize}
\item [3.5] - I’m interested in the repository for my own needs
\end{itemize}

The average of self-use answers was 7.86. 
56\% of the participants reported high self-use motivation, 61\% of GitHub participants and 43\% of paid ones.
Note that while usually the probabilities in the entire population, in GitHub, and among paid participants are rather similar, in this case the probabilities are quite different.
`Scratching your own itch' is a well known motivation in open-source \cite{Raymond:1999:CB:580808} so one would expect a higher probability in GitHub.
On the other hand, many companies produce organizational software that does not have personal uses, so 43\% of paid participants which self-use may sound rather high.
\hide{
\begin{figure}[!ht]
\centering
\includegraphics[width=0.88\textwidth,trim={0mm 0mm 0 0mm},scale=0.8]{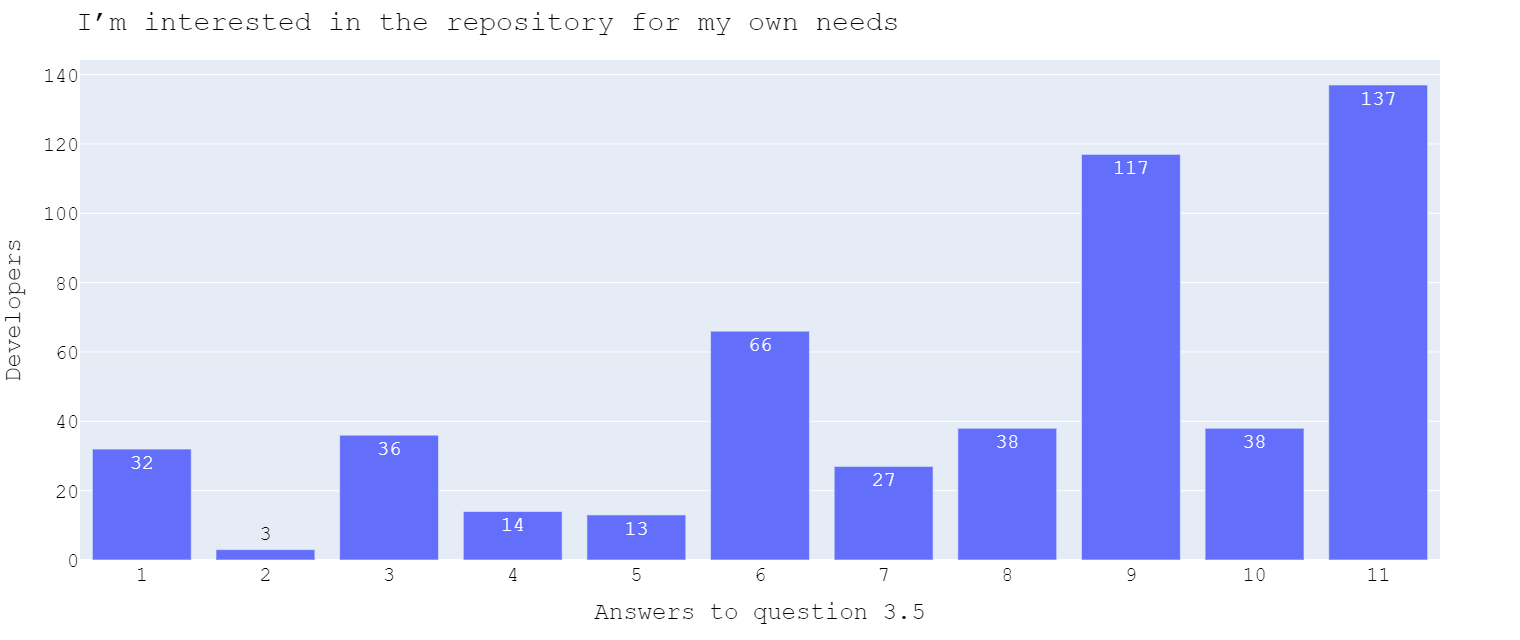}
\caption{\label{fig:self_use}
Answers distribution of self-use question.
}
\end{figure}
}
The correlation of self-use with motivation is 0.16.
Self-use has a recall of almost two thirds, 0.65.
This seems to be a unique attribute of open-source, enabling people to develop the software that they need.
The precision is 0.61 and the precision lift is 0.17, moderate values.
This might be since people see satisfying their need as a task to complete and not an enjoyable activity.
Indeed, self-use and enjoyment have a Pearson correlation of just 0.16.

\subsubsection{Ideology}\label{sect:ideology}

Ideology \cite{gerosa2021shifting, belenzon2008motivation} is based on the question: 
\begin{itemize}
\item [2.18] - I contribute to open source due to ideology
\end{itemize}

53\% of the participants reported high ideology-based motivation, rising to 61\% of GitHub participants, aligned with the ideological roots of open-source development \cite{belenzon2008motivation}.
49\% of paid participants also gave high answers regarding contribution due to ideology.
That can be either due to many people being paid to contribute to open-source, or a common habit of paid developers to contribute to open source in their free time.
Regardless of the reason, the popularity is surprisingly high.
But we note that in the answers to the open question participants said that different ideologies (e.g., `Software should be free' \cite{stallman2007software}, `Social good') can lead to contribution to open-source software, and that a finer distinction is needed.
\hide{
\begin{figure}[!ht]
\centering
\includegraphics[width=0.88\textwidth,trim={0mm 0mm 0 0mm},scale=0.8]{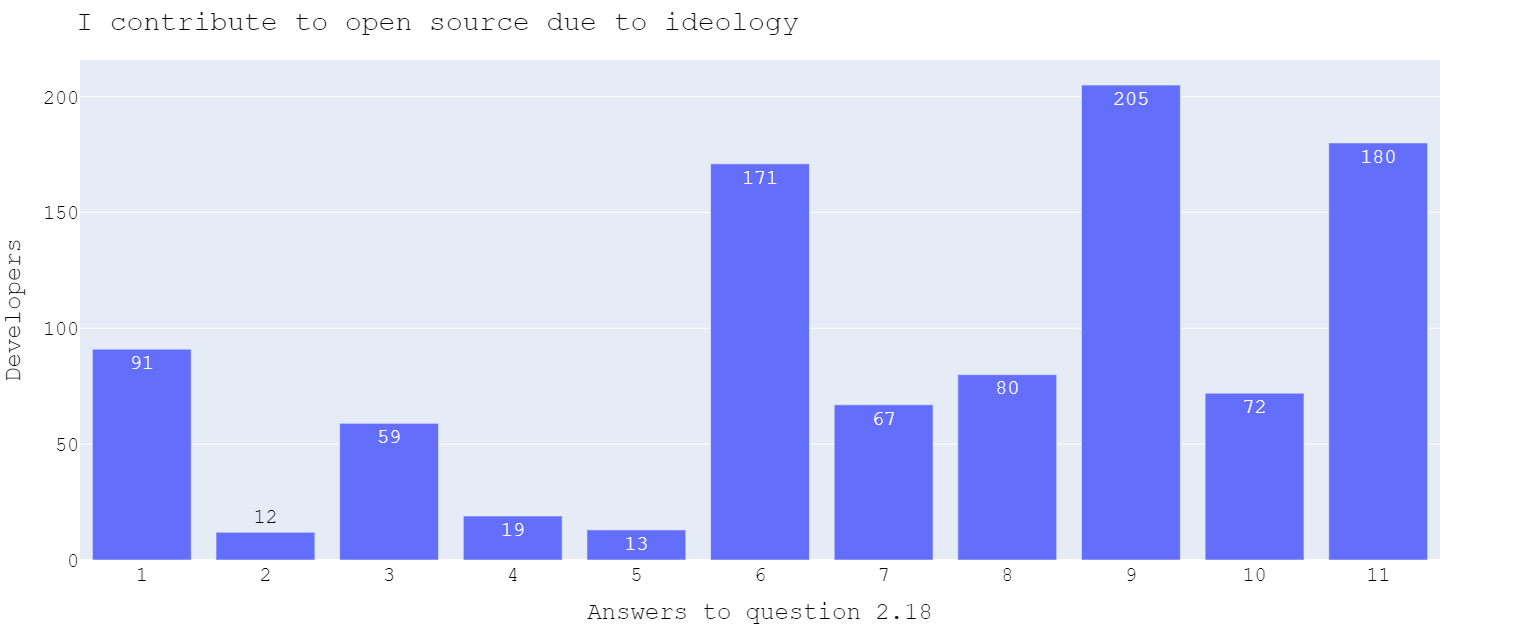}
\caption{\label{fig:2_18_ideology}
Answers distribution of ideology question.
}
\end{figure}
}
The average of ideology answers was 7.34.
The correlation of ideology with motivation is 0.14, the lowest other than for hostility.
The precision of ideology is 0.59, the precision list is 0.13, and the recall 0.60.

\subsubsection{Recognition}\label{sect:recognition}

We measure recognition \cite{DBLP:journals/infsof/BeechamBHRS08, gerosa2021shifting, riehle2007economic, Raymond:1999:CB:580808, VonKrogh:2012:CRM:2481639.2481654} by the following questions:

\begin{itemize}
\item [2.13] I contribute to open source in order to have an online portfolio 
\item [2.14] I try to write high quality code because others will see it
\item [3.15] I get recognition due to my contribution to the repository
\item [3.24] In the past year, members of my GitHub community asked questions that show their understanding of my contributions (based on \cite{Kluger17FLS})
\item [3.25] In the past year, members of my GitHub community expressed interest in my contributions (based on \cite{Kluger17FLS})
\end{itemize}

The average of all the questions was 7.33, lower than all positive motivators. 
48\% of the participants reported high recognition-based motivation, 49\% of the GitHub ones, and 52\% of paid ones.
The correlation of recognition with motivation is 0.27.
\hide{
\begin{figure}[!ht]
\centering
\includegraphics[width=0.88\textwidth,trim={0mm 0mm 0 0mm},scale=0.8]{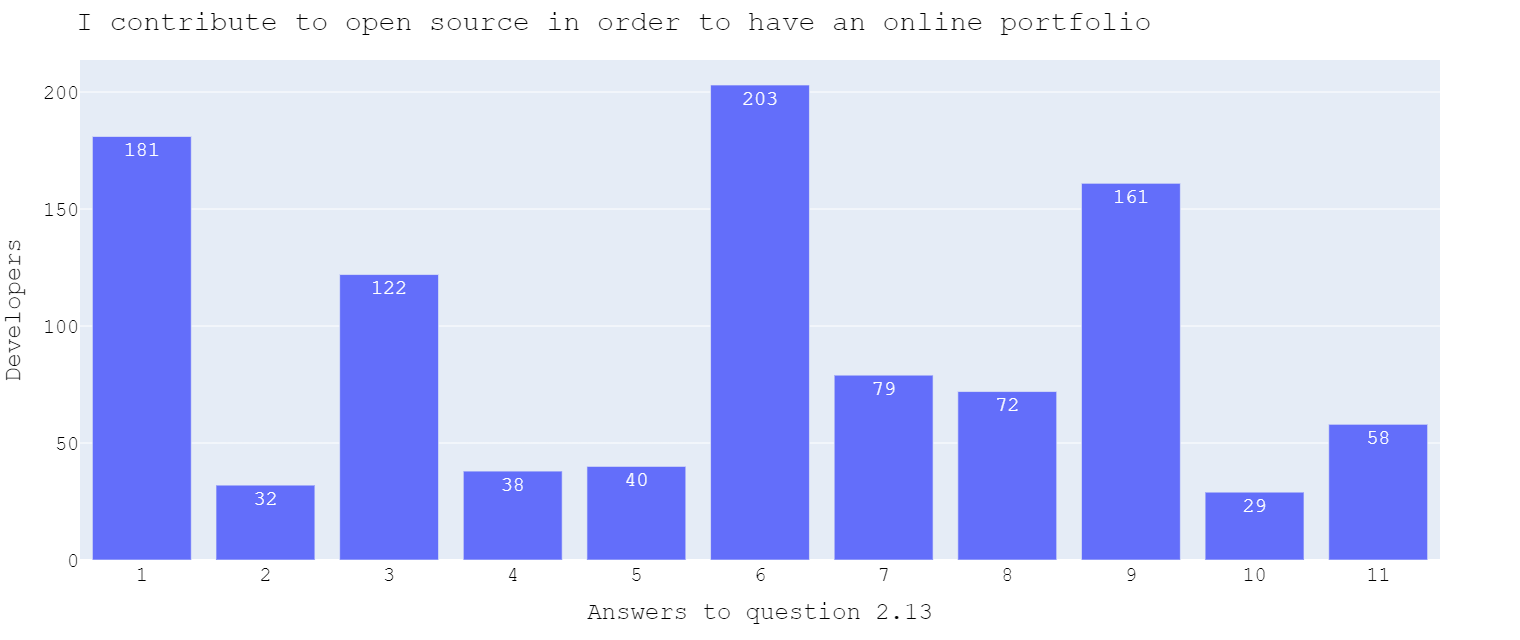}
\caption{\label{fig:2_13_online_protfolio}
Answers distribution of recognition question.
}
\end{figure}
}
The precision is 0.60 and the precision lift 0.18, indicating a boost to motivation.
The recall is 0.56, moderately higher than the recognition hit rate.

\subsubsection{Payment}\label{sect:payment}

Payment \cite{DBLP:journals/infsof/BeechamBHRS08, gerosa2021shifting, benabou2003intrinsic, riehle2007economic} is based on the yes/no question:
\begin{itemize}
\item [3.c] - I'm being paid for my work in this repository
\end{itemize}

We note that remuneration in open-source projects may have many facets.
Developers may accrue income from donations or lectures.
Their work on the project may help them secure future positions or gain access to future consulting contracts.
In the interest of simplicity and precision we use the objective criterion of specifically being paid a salary to define payment.

41\% of the participants that answered the payment question said that they are paid.
38\% of participants of GitHub projects that answered the payment question said that they are paid.

Payment has precision of 0.58 and precision lift of 0.10, making it one of the weakest motivators in general and the weakest for its hit rate.
Payment is also the only motivator that had a negative precision lift when we run the same analysis on the follow-up survey.

Its recall is 0.49, less than the positive rate of high motivation which is 0.52.
The correlation of payment with motivation is 0.15, lowest than all but ideology and hostility.

\subsubsection{Community}\label{sect:community}

We measure community \cite{DBLP:journals/infsof/BeechamBHRS08, gerosa2021shifting, MCCLELLAND1961Achieving, beggan1992social,Asproni2004,Ye:2003:TUM:776816.776867,10.1145/3422392.3422433} 
by the following questions (which we asked to answer only if you are not the only developer in the project):

\begin{itemize}
\item [3.13] Belonging to the community is motivating my work on the repository 
\item [3.14] The community is very professional
\item [3.20] The repository’s community of developers is more motivated than that of other repositories
\item [3.24] In the past year, members of my GitHub community asked questions that show their understanding of my contributions (based on \cite{Kluger17FLS})
\item [3.25] In the past year, members of my GitHub community expressed interest in my contributions (based on \cite{Kluger17FLS})
\end{itemize}

Note the questions 3.24 and 3.25 are about recognition from the community and therefore appear in both motivators.

The average of community answers was 7.36. 
40\% of the participants reported high community-based motivation, lower than all positive motivators.
This is based on 40\% of the GitHub participants and 44\% of the paid ones.
The correlation of community with motivation is 0.42, the second highest.
\hide{
\begin{figure}[!ht]
\centering
\includegraphics[width=0.88\textwidth,trim={0mm 0mm 0 0mm},scale=0.8]{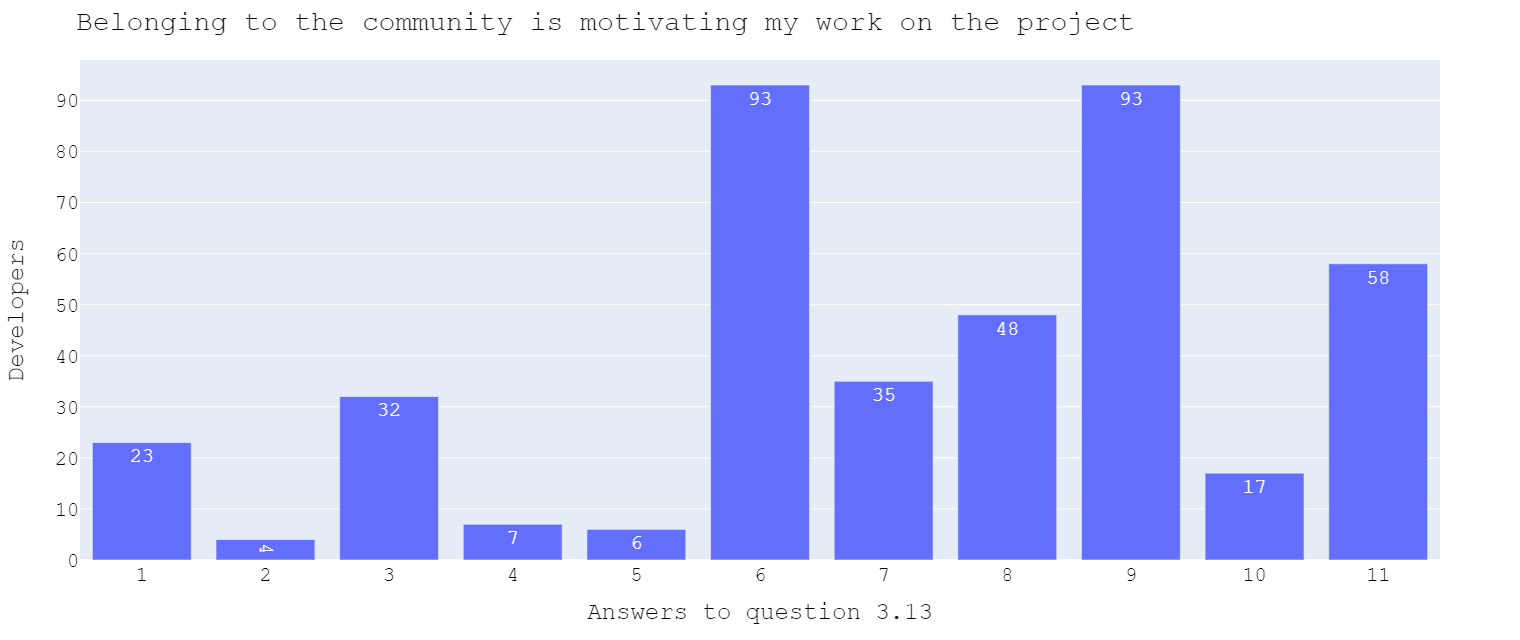}
\caption{\label{fig:3_13_belonging}
Answers distribution of community question.
}
\end{figure}
}%
The precision is 0.67 and the precision lift is 0.35, higher than all other motivators.
Hence, though the community motivator is not common, when it exists, the probability of high motivation is higher.
The recall is 53\%, not very high yet 29\% higher than the hit rate.

\subsubsection{Hostility}\label{sect:hostility}

\begin{figure}[!ht]
\centering
\includegraphics[width=1.0\textwidth,trim={0mm 0mm 0 0mm},scale=0.3]{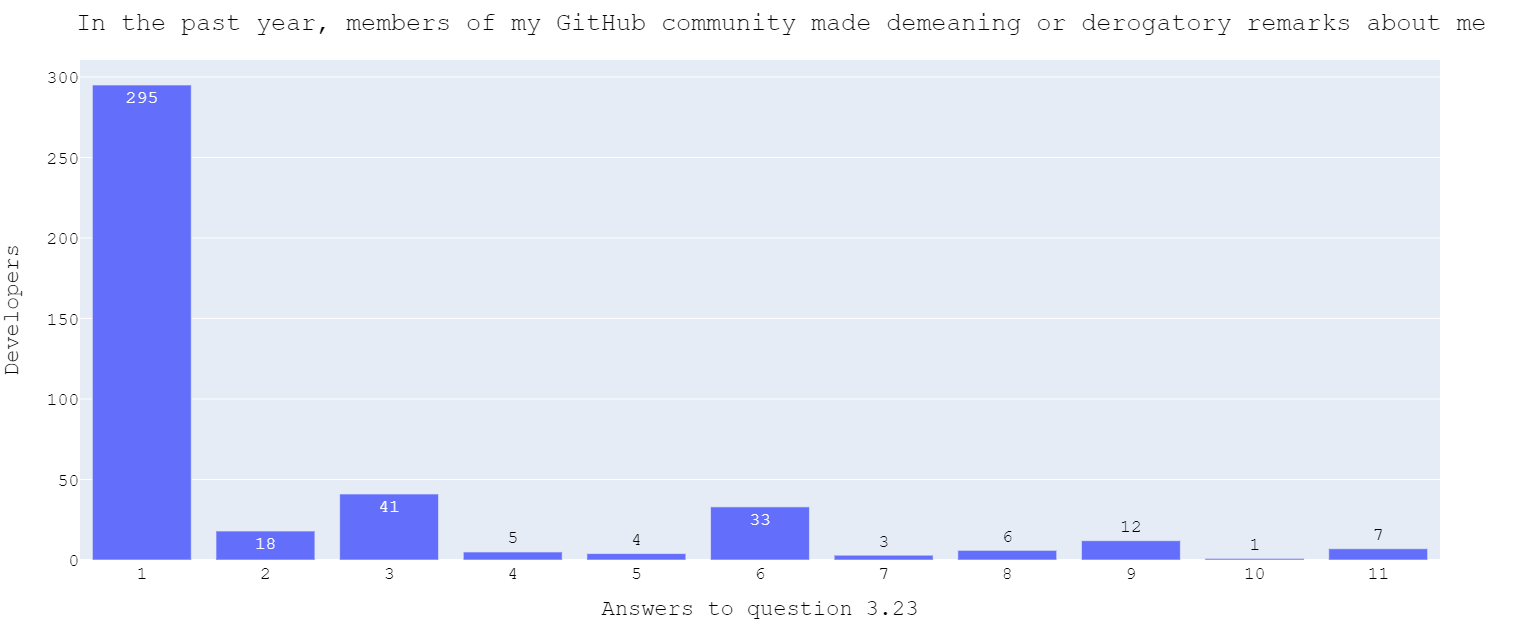}
\caption{\label{fig:q96.1_derogatory} Answers distribution of a hostility question.
}
\end{figure}

Hostility can be viewed as a community with negative influence.
Hostility hurts motivation hence it is a demotivator and not a positive motivator.
We measure hostility \cite{Raman2020StressAB, Adams1963TOWARDAU, Herzberg1959Motivation, Cortina2001Incivility} by the following questions (which we asked to answer only if you are not the only developer in the project):

\begin{itemize}
\item [3.6] We have many heated arguments in the community
\item [3.7] I wish that certain developers in the repository will leave
\item [3.22] In the past year, members of my GitHub community put me down or were condescending to me (based on \cite{Cortina2001Incivility})
\item [3.23] In the past year, members of my GitHub community made demeaning or derogatory remarks about me (based on \cite{Cortina2001Incivility}, Figure \ref{fig:q96.1_derogatory})
\end{itemize}

The average of hostility answers was 2.80. 
Only 7\% of the participants reported high hostility, 4\% of the GitHub participants and 7\% of the paid ones.
The recall is just 8\%, yet it is higher than the hit rate.
The correlation of hostility with motivation is 0.01.
This is the lowest correlation, close to zero yet not the large negative correlation which is expected.

Surprisingly, hostility also has a relatively high precision of 0.65 and a high precision lift of 0.30.
One would expect that knowing that someone suffers from hostility will \emph{reduce} the probability of high motivation, instead of the increase that we see.
But the participants who reported high hostility also reported higher averages for all motivators besides payment.
A possible explanation is that those participants kept contributing to the project due to the other motivators; those who suffered from hostility and did not have other reasons to stay probably left.
Note, however, that we had only 9 people that reported both high hostility and high motivation, so the analysis is not robust.

We identified 10 pairs of developers which contribute to the same project.
This allowed us to evaluate their agreement on hostility.
Surprisingly, when a person reports heated arguments (question 3.6), the probability that the other participant will agree is just 50\%.
For the rest of the hostility questions, the other participants never claimed high hostility too.
For comparison, in importance and challenge, which also describe aspects of the project, if one participant provided a high half answer (6 or above), the other always agreed.
This provides an important indication that hostility might go unnoticed.

\subsection{Relations Between Motivators}

The above motivators are not all independent.
It is therefore interesting to see how they correlate with each other.
We now consider motivation and all its motivators as variables and examine the relations between them.
We calculated the Pearson correlation between every pair of variables, and looked for connected components on the variables graph, in which an edge exists given a high Pearson correlation.
 
 We use two thresholds of 0.8 and 0.5 Pearson correlations.
The aim of the thresholds is to identify strong relations and moderate relations.
We compute the correlations of the variables using three populations: all participants, 284 paid participants (as in Section \ref{sect:payment}), and 377 GitHub participants (those who reported a project hosted on GitHub).
For each pair of variables, we use the answers of all participants that answered at least one question per variable.
The use of the different populations is to see if there are differences in motivation relations in different contexts.

\begin{figure}[!ht]
\centering
\includegraphics[width=1.15\textwidth]{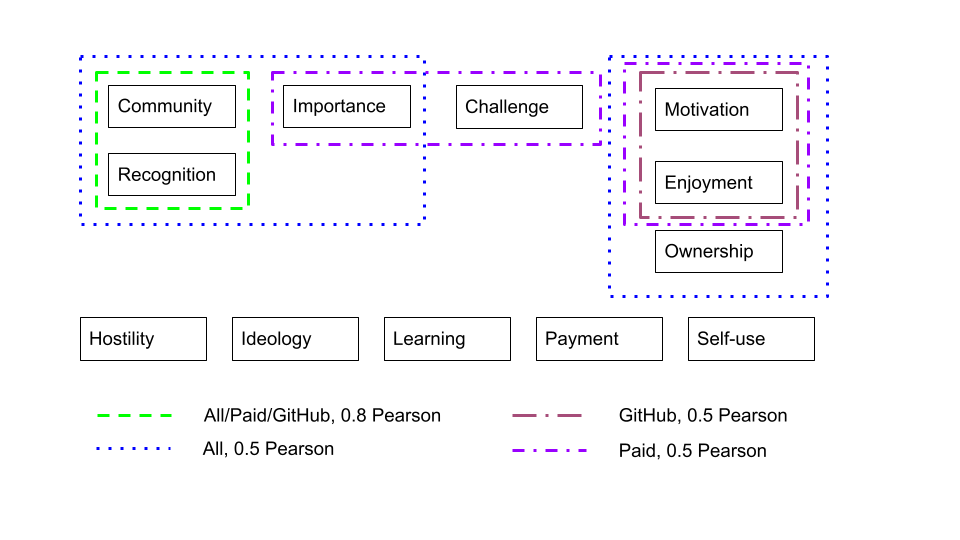}
\caption{\label{fig:factors_relations}
Relations between variables at Pearson levels of 0.8 and 0.5, on the whole population, paid participants, and GitHub participants.
}
\end{figure}

Figure \ref{fig:factors_relations} presents the sets of correlated variables identified in different contexts.
In all contexts, the 5 variables at the bottom have correlation lower than 0.50.
In all 3 populations `Recognition' and `Community' are strongly correlated.
However, as noted above, recognition and community share two questions regarding recognition from the community (questions 3.24 and 3.25).
Removing the common questions, the Pearson correlation of community and recognition is 0.37 on the entire population, 0.35 on GitHub participants, and 0.26 on paid participants.
Hence, their actual correlation is much lower.

`Enjoyment' and `Motivation' are moderately correlated for GitHub users and for paid users, and also correlated with `Ownership' on all participants.
`Importance' is moderately correlated with `Recognition' and `Community' on all participants and with `Challenge' on paid participants.

In conclusion, the motivator with the highest correlation with motivation is enjoyment, as discussed in Section \ref{sect:enjoyment}.
Other motivators are less correlated, implying that they are not redundant and each one exposes a different behavioral aspect.

\subsection{Motivation Improvement Analysis}\label{sect:motivation-improvment-analysis}

An important goal in many data analyses is to uncover causal relations.
But causality is hard to define rigorously because it is hard to ascertain that motivator $A$ caused outcome $B$.
The usual approach is to look at correlations between motivators and outcomes in a given dataset, as we did in Section \ref{sect:motivating-factors}.
We now extend this to look at the \emph{dynamics across time}: the possible correlation between \emph{a change} in a motivator and \emph{a change} in the outcome.
For example, we want to see whether an increase in the sense of ownership of a project predicts an increase in motivation.

Such ``co-change'' analysis \cite{Amit2021CCP} is important for the following reason.
If causality exists, meaning that in certain contexts motivator $A$ causes outcome $B$, then a change in $A$ will cause a change in $B$.
But co-change of two motivators does not necessarily imply causality.
By identifying instances of co-change, where a change in $A$ correlates with a change in $B$, we therefore identify cases where causality may be at work.

Note, however, that the actual relationship between motivators and general motivation may be conditioned on other motivators.
In this subsection we look at the co-change of motivators and general motivation alone, regardless of context.
In the next subsection we consider all the motivators together, to handle cases where the effect of $A$ on $B$ is conditioned on another motivator $C$.

The change data comes from comparing the original survey and the follow-up survey.
In the original survey \devEmailNum developers provided their emails.
A year after the last response, we reached out and asked them to answer the survey again.
We asked them to answer on the same project if they are still active in it.
This allowed us to compare the answers of the same person over time.
We had \followupParticipants follow-up participants in total.
60 of them continued in the same project, and these are the ones we analyze here.
For each of them, we look at increases in the motivators and general motivation from one year to the next.
Note that if a person reported 3 for ownership in the first survey, and 4 in the follow-up, this is an increase regardless of the values being low. 

\begin {table}[h!]\centering
\caption{ \label{tab:steps-motivation-factor} Motivation Improvement Over Time Predictability by Motivator  }
\begin{tabular}{ | l| c| c| c| c| c| }
\hline
Motivator & Improvement & \multicolumn{4}{c|}{Prediction of improved motivation} \\
\cline{3-6}
     &     rate    & Accuracy & Precision & Prec.\ lift & Recall \\
\hline
Challenge & 0.33 & 0.53 & 0.10 & -0.50 & 0.17\\ \hline
Ideology & 0.30 & 0.60 & 0.17 & -0.17 & 0.25\\ \hline
Importance & 0.30 & 0.70 & 0.33 & 0.67 & 0.50\\ \hline
Learning & 0.30 & 0.70 & 0.33 & 0.67 & 0.50\\ \hline
Enjoyment & 0.28 & 0.71 & 0.34 & 0.71 & 0.48\\ \hline
Recognition & 0.27 & 0.72 & 0.39 & 0.95 & 0.47\\ \hline
Self-use & 0.22 & 0.78 & 0.46 & 1.31 & 0.50\\ \hline
Ownership & 0.17 & 0.77 & 0.45 & 1.27 & 0.35\\ \hline
Hostility & 0.16 & 0.70 & 0.20 & -0.01 & 0.15\\ \hline
Community & 0.16 & 0.75 & 0.39 & 0.93 & 0.28\\ \hline
\end{tabular}
\end{table}

\hide{
\begin {table}[h!]\centering
\caption{ \label{tab:down-steps-motivation-factor} Downwards Person Change Over Time Predictability by Motivator }
\begin{tabular}{ | l| l| l| l| l| l  | }
\hline
Motivator & Improvement Rate & Accuracy & Precision & Precision Lift & Recall \\
\hline
Self-use & 0.43 & 0.68 & 0.77 & 0.40 & 0.61\\ \hline
Ideology & 0.37 & 0.58 & 0.68 & 0.24 & 0.45\\ \hline
Enjoyment & 0.36 & 0.56 & 0.66 & 0.19 & 0.43\\ \hline
Importance & 0.35 & 0.60 & 0.71 & 0.30 & 0.45\\ \hline
Recognition & 0.34 & 0.57 & 0.69 & 0.25 & 0.41\\ \hline
Challenge & 0.33 & 0.55 & 0.65 & 0.18 & 0.39\\ \hline
Community & 0.28 & 0.50 & 0.59 & 0.07 & 0.30\\ \hline
Learning & 0.28 & 0.50 & 0.59 & 0.07 & 0.30\\ \hline
Ownership & 0.27 & 0.54 & 0.67 & 0.21 & 0.33\\ \hline
Hostility & 0.12 & 0.46 & 0.55 & 0.00 & 0.12\\ \hline
Payment & 0.10 & 0.48 & 0.67 & 0.21 & 0.12\\ \hline
\end{tabular}
\end{table}
}

The probability of improvement in general motivation, i.e.\ the positive rate, was 20\%.
Since we included only developers that stayed in the same project for at least a year, probability in the whole population is probably even lower.

The probability of improvement in the different motivators is at most 33\% (Table \ref{tab:steps-motivation-factor}).
When the precision lift is positive, it tends to be very high.
We could not find out why the lift is negative for challenge and ideology.
One could expect a larger negative lift for hostility, which did not materialize.
This may be explained by developers having other motivators that offset hostility as explained in Section \ref{sect:hostility}.

The recall is up to 50\% for many motivators and higher than their improvement rate (hit rate).
This shows that improvement in importance, learning, enjoyment, recognition, and self-use are common when motivation improves.

Only a single person that was not originally paid received a payment in the follow-up, therefore we did not apply co-change analysis to this motivator.

Co-change analysis can be performed in the downward direction too: given a decrease in a motivator, how common is a decrease in general motivation.
Results are quite similar to the upward direction and given in the supplementary materials. 

The followup survey can also be considered a replication of the original survey.
We used all \followupParticipants participants that answered the follow-up survey to run the analysis of predicting motivation (as in Table \ref{tab:high-answer-motivation-factor}).
We found a positive precision lift for all motivators besides payment.
The agreement supports the results in general.
The disagreement in payment indicates that that result is not robust.

\hide{
If we had a minimal model of perfect accuracy predicting changes in general motivation given changes in motivators, we could predict causality.
A minimal model means that no variable can be discarded, since the variable influences the prediction, hence causing a change.
Perfect accuracy means that no other causes exist since the existence of such a motivator would have led to changes that we could not predict.
}

\subsection{A Combined Motivation Model}\label{sect:model}

All the motivators have a positive precision lift of at least 10\% (Table \ref{tab:high-answer-motivation-factor}).
This is aligned with the prior work claiming their positive influence.
However, the highest precision lift is just 35\%, with 67\% precision, for the `Community' motivator.
This means that none of the motivators is a sufficient condition for high motivation or close to it.
Hence, a combination of motivators is needed to reach high motivation.

The inputs of machine learning models are named ``features''.
So far, we investigated each motivator as a single feature, ignoring all other motivators.
This type of analysis suffers from the threat of confounding variables on one hand, and does not leverage the full power of the data on the other hand.
We therefore built combined models to predict motivation, based on all the motivators as features.

The predicted concept was high motivation, operationalized as before by answers of 9 (`somewhat agree') or above to ``I regularly have a high level of motivation to contribute to the repository''.
We had 345 participants that answered this question.
The positive rate is 52\%.

We used the scikit-learn package for classification algorithms \cite{scikit-learn}.
We used low-capacity small models such as decision trees and logistic regression in order to obtain simple interpretable models which are also rather robust to overfitting \cite{arpit2017closer}.
We also used models of moderate capacity such as random forests, boosting, and neural networks to build models of better representation ability and performance.
In order not to bias our results, 70\% of the dataset was used for training and performance was evaluated on the 30\% remaining test data.

The performance of all the models was rather close, with accuracy ranging from 62\% to 77\%.
Amusingly, the highest accuracy on the test set (77\%) was reached by a single node tree, checking high enjoyment.
As Table \ref{tab:high-answer-motivation-factor} showed, the accuracy when using enjoyment on the whole dataset is just 64\%, so this result is accidental.
The model with the second highest accuracy (76\%) was a neural network \cite{lecun2015deep}, whose capacity is high.
The simple model of highest accuracy was a logistic regression model \cite{verhulst1845resherches} which reached accuracy of 72\%.
Its intercept was -2.04, indicating a general tendency for low motivation.
Hostility had a strong negative coefficient of -0.46.
All the other motivators had positive coefficients.
The highest were enjoyment with 1.13, self-use with 0.61, and importance with 0.59.

Note that models can assign different weights to false positives and false negatives, and trade off precision and recall.
Using this, we could build a precision-favoring decision tree model \cite{quinlan2014c4} with precision of 81\% and recall of 41\%.
Conversely, a recall-favoring stochastic gradient descent (SGD) model \cite{10.1145/775047.775151} reached recall of 96\% with precision of 62\%. 

We also modeled the co-change dataset of Section \ref{sect:motivation-improvment-analysis}, to predict a change in the motivation based on changes in the motivators as features.
Such a model is of interest since assuming that motivation is a function, a co-change model can predict the result of a change.

Two properties of such modeling deserve special attention: accuracy and minimality.
We first discuss accuracy.
A co-change model allows us to predict \emph{the motivation change} given any \emph{motivator's change}.
Perfect accuracy assures us that there are no other external causal variables influencing the samples in our dataset.
Assume by contradiction such a variable $c$, other than the model variables.
Hence there is a behavior function $g$ and an assignment of values such that $g(c_{1}, v_{1},...v_{n}) \neq g(c_{2}, v_{1},...v_{n})$ where $ v_{1},...v_{n}$ are the values of the model variables.
However, since we have perfect prediction given the model variables, it should be that
$g(c_{1}, v_{1},...v_{n}) = m(v_{1},...v_{n}) = g(c_{2}, v_{1},...v_{n})$ --- a contradiction.
Hence such a variable cannot exist.
Perfect accuracy is rare, and mostly indicates a problem in the analysis and not capturing all causal variables.
However, the accuracy bounds the influence of such external variables.

As for minimality, consider decision trees \cite{quinlan2014c4} as models.
For each leaf, variables that do not appear in the path to this leaf do not influence the prediction.
On the other hand, each variable along the path is necessary, and a change in its value will change the prediction.
In this sense, the model is minimal and every variable along the path is required.
All the variables that we use here are mutable (can change, in contrast for example to the project creation year).
With both perfect accuracy and minimality, each change is explained by the model and the removal of any variable will hurt the prediction of some changes.

\begin{figure}[!ht]
\centering
\includegraphics[width=0.88\textwidth,trim={0mm 40mm 0 40mm},scale=0.8]{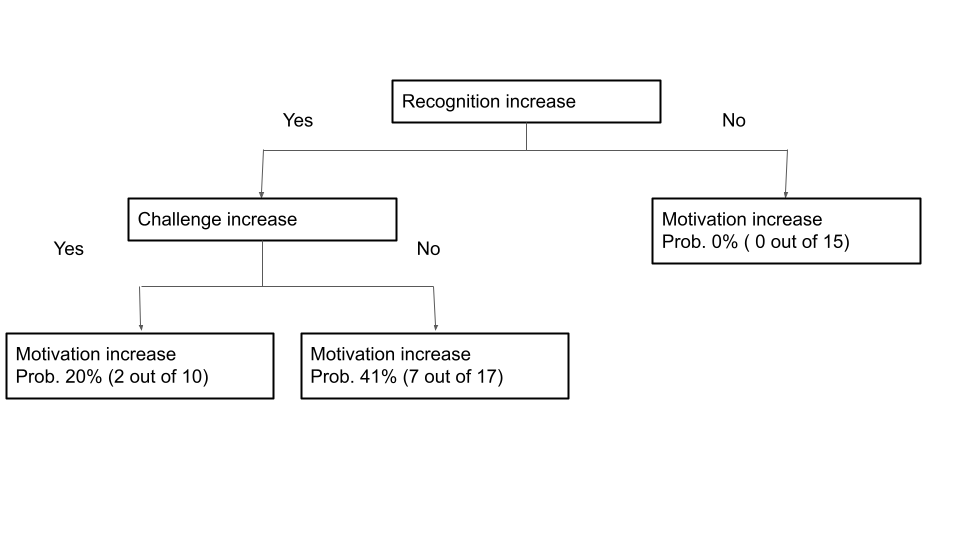}
\caption{\label{fig:co-change-model}
A decision tree predicting motivation improvement}
\end{figure}

When aiming for high accuracy, we built models based on AdaBoost \cite{freund1997decision} and Neural Networks \cite{lecun2015deep} which reached accuracy of 94\%.
Aiming for either precision or recall, 100\% were reached.
The size of these models is high and far from minimality, which is the price for achieving the high accuracy.
Alternatively, we found a small model, presented in Figure \ref{fig:co-change-model}, whose ``recognition increase and challenge decrease'' leaf reaches 78\% recall with 41\% precision.
Note, however, that the dataset very small dataset with a high VC dimension \cite{vapnik1971VC, vapnik2013nature} (due to having many question and wide scale), and therefore the threat of noise is very high.
Also, the dataset probably does not fully represent motivation complexity.
A larger dataset will probably better represent human motivation behavior but will require a larger model and have lower performance.

\section{Analysis of Validity and Reliability}

In the previous analysis we analyzed the data as if it is completely reliable.
However, the reliability might be limited in many ways.
Since the data is given, what we do in this section is to evaluate its reliability from various aspects.

When using the answers of participants, one should check which population they represent.
We compare our survey demographics to the demographics of the Stack Overflow survey, answered by around 80 thousand developers world-wide (Section \ref{sect:stackoverflow}).
We used two channels in order to reach participants: direct emails to developers contributing at GitHub and social media.
In Section \ref{sect:population-difference} we use machine learning to see how different these populations are. 
We grouped questions into motivators based on their content, regardless of the answers to them.
In Section \ref{sect:factor-coherence} we examine the coherence of the motivators and compare them to grouping based on the answers.
The follow-up survey allows us to evaluate the stability of answers, comparing a person's answer in two different dates (Section \ref{sect:follow-up-stability}).
It also provides an additional dataset on which we can check the degree in which our results reproduce.
Last but not least, we investigate reliability in the answers themselves, from typos to mistakes and biases (Section \ref{sect:answer-validity}).

\subsection{Comparison to the Stack Overflow Survey Demographics}\label{sect:stackoverflow}

It is hard to define `developer' (e.g., by education, profession, minimal activity) and therefore to define a representative developers population.
However, a possible comparison is provided by the  `\href{https://stackoverflow.com/}{Stack Overflow}' (SO) annual developer survey.
SO is a leading questions and answers website for programming, with more than 100 million monthly visitors.
The 2021 survey, overlapping our survey period, was answered by over 80 thousand developers \cite{Stackoverflow2021Survey}.
As far as we know, this is the largest survey of developers and therefore an interesting comparison.

\hide{Our survey participants come from 74 countries, compared to 180 countries in SO.
In our survey the leading countries are the United States, Germany, Great Britain, India, and France.
These are the same leading countries as in SO, but in SO India is in second place and the Western countries have a somewhat lower representation.
}
4.2\% of our participants identified as females, 95.1\% as males, and the rest as others.
This is close to the ratio in the 2021 SO survey, which was 5.3\%.
It is also close to ratios of 4.8\% \cite{terrell2017gender} and 11.2\% \cite{10.1145/2597073.2597129} in other sources.

\hide{
Figure \ref{fig:age_dist} presents the age distribution.
The distributions are rather close with slightly higher ages in our survey.

\begin{figure}[!ht]
\centering
\includegraphics[width=1.0\textwidth,trim={0mm 0mm 0 0mm},scale=0.3]{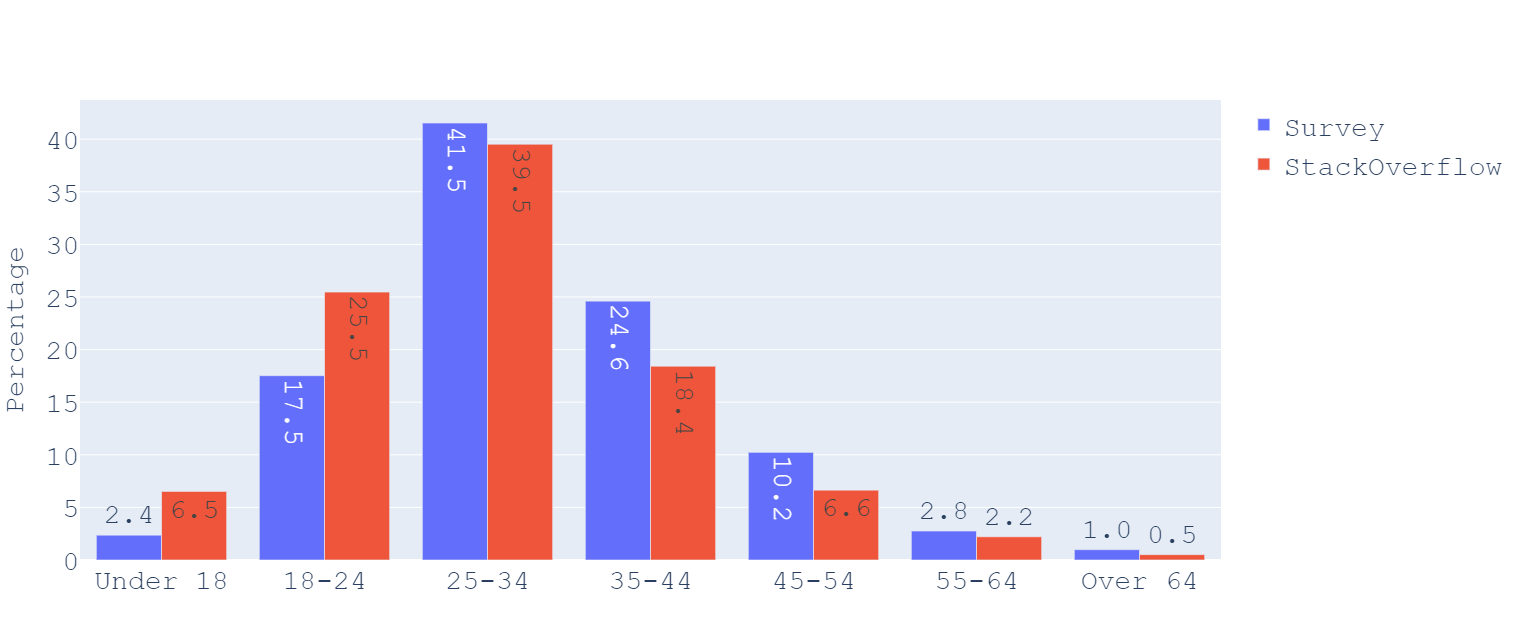}
\caption{\label{fig:age_dist} Age Distribution
}
\end{figure}
}

\begin{figure}[!ht]
\centering
\includegraphics[width=1.0\textwidth,trim={0mm 0mm 0 0mm},scale=0.3]{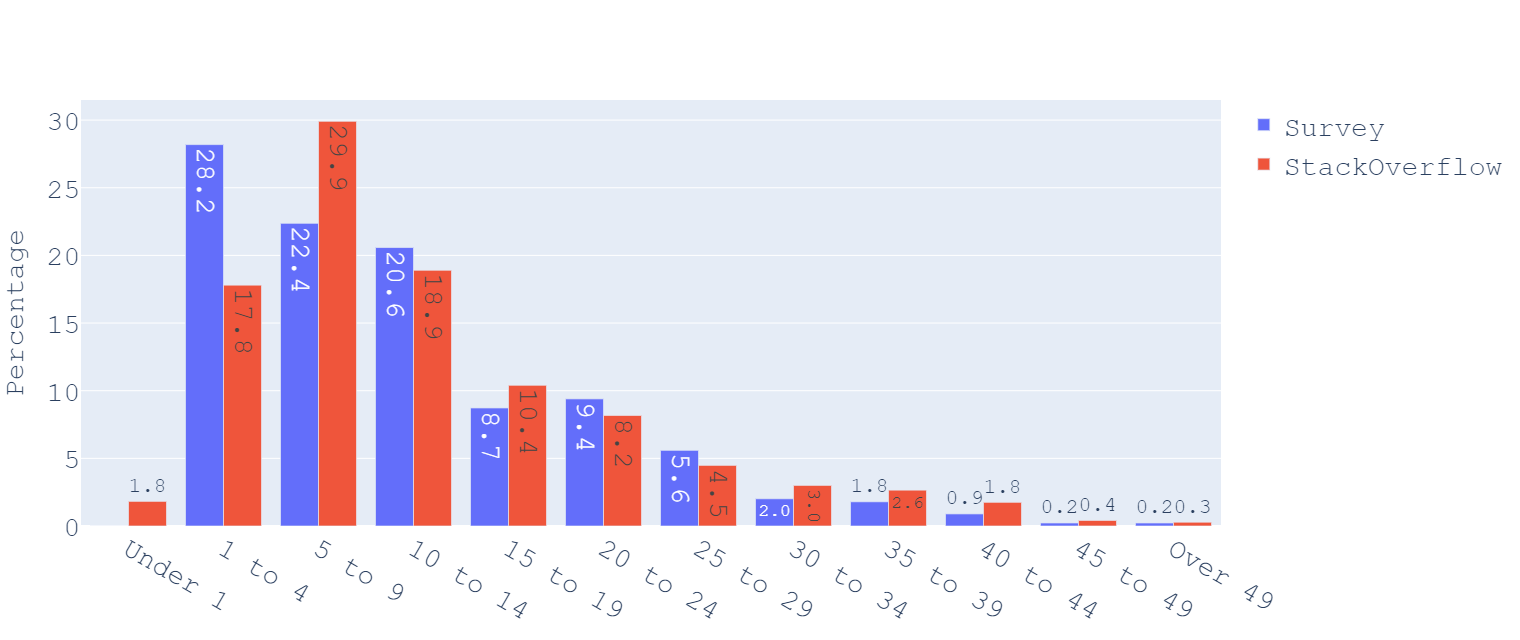}
\caption{\label{fig:Year_of_exp_Dist} Years of Experience Distribution
}
\end{figure}

80.7\% of the participants in our survey work as professional programmers, compared to 69.7\% in SO.
The average years of programming experience in our survey (professional or not) is 11.1, representing very experienced developers.
Figure \ref{fig:Year_of_exp_Dist} shows that in both surveys about half of the participants have at least 10 years of experience.

\begin{figure}[!ht]
\centering
\includegraphics[width=1.0\textwidth,trim={0mm 0mm 0 0mm},scale=0.3]{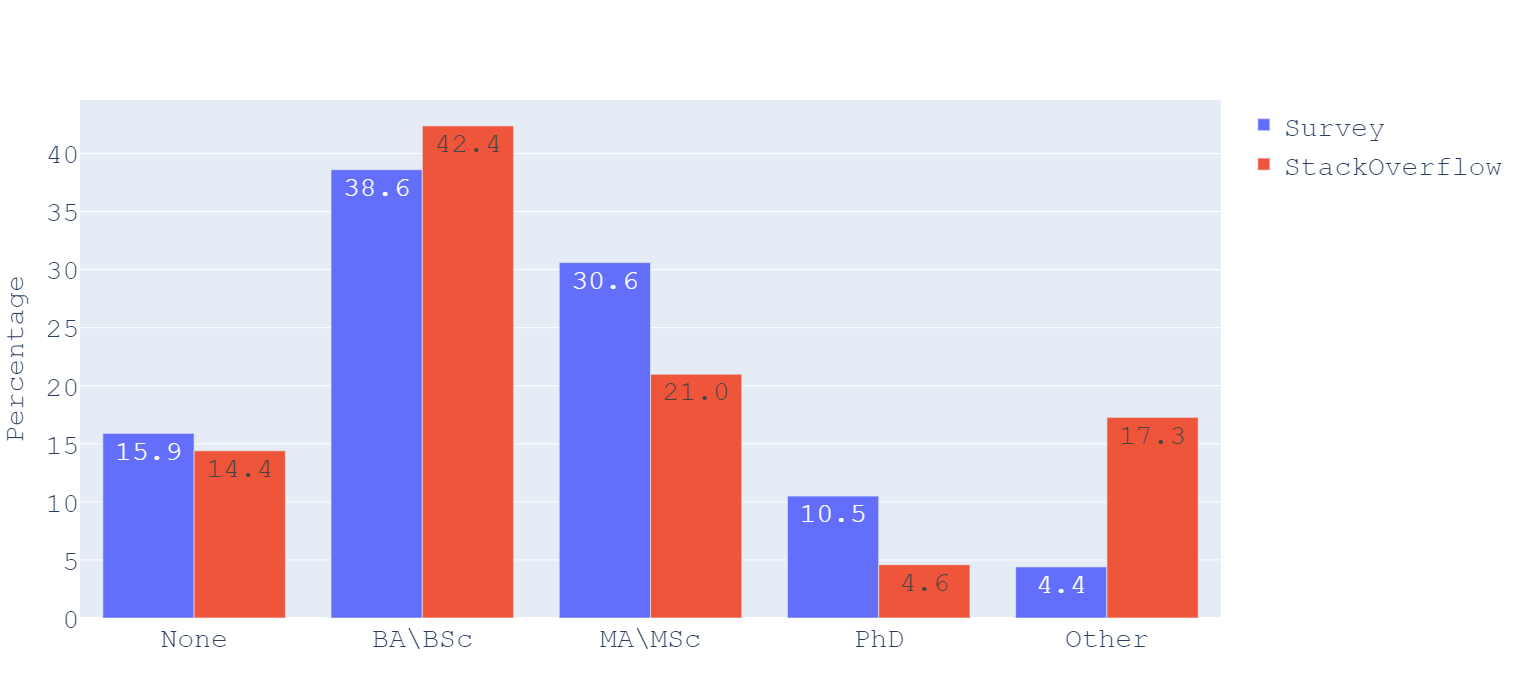}
\caption{\label{fig:Academic_Dist} Academic Background
}
\end{figure}

Our survey had more participants with high degrees than SO, as shown in Figure \ref{fig:Academic_Dist}.
The academic domains in our survey were: computer science 46.3\%, technology 32.0\%, science 9\%, business 3.4\%, math 2.5\%, arts 2.1\%, and the rest from other domains.

To conclude, the demographics in our survey are close but not the same as those in Stack Overflow.
For example, more people identify as professional developers in our survey.
But it seems that the threat of not representing the developers community is low.

\subsection{Differences Between GitHub and Social Media Participants}\label{sect:population-difference}

We received answers from two different populations.
20\% of the participants were developers contributing to a GitHub project, which were recruited via direct emails.
The other 80\% were reached out to in convenience sampling \cite{etikan2016comparison, acharya2013sampling}, using messages in social media.
Differences between the populations might lead to investigating two different behaviors as an averaged one.
It is common to compare populations using a comparison of distributions of demographic variables (e.g., age, gender).
However, the relevant questions appeared at the end of our survey, and many developers that did not contribute to open source did not reach this part in our first phase.
Instead, we reduced the problem of populations difference into a supervised learning problem, trying to predict the source of the participants using the questions in the first part of the survey.
The features included open-source specific questions such as `I contribute to open source in order to become a better programmer' and `I contribute to open source due to ideology'.
Nevertheless, a decision tree model, suitable to a small number of samples, reached an accuracy of only 78\%.
Even high-capacity models such as SVM or Neural Networks reached an accuracy of only 80\%.
Note that since the positive rate is 20\%, the majority rule betting that all the developers are from social media would also lead to an accuracy of 80\%.
Such low predictive power does not mean that the populations are similar.
However, it means that there is no big obvious difference, even when considering contribution to open source, based on the personal questions.
Therefore, we can analyze both populations together, getting a larger dataset and leading to more robust results.

\hide{Though, our response rate of GitHub developers was rather low, raising the threat of response bias \cite{gove1977response,mazor2002demonstration} and therefore our GitHub developer dataset might not represent the GitHub developers population.
Due to that we cannot claim that our entire sample represents their population too.
}

\subsection{Internal Coherence of Motivators}\label{sect:factor-coherence}

The motivators were represented in the survey by one or more questions each.
The use of multiple questions (e.g.\ for `community') allows us to treat them as labeling functions of the same concept and evaluate their agreement \cite{NIPS2016_6523, archimedes}.
The agreement, measured by the average Pearson correlation of the related questions, reflects the internal coherence of these motivators.
Low coherence might be due to our subjective grouping of questions or due to human nature.

\hide{
Some of the motivators were represented by more than one question (e.g.\ `community').
These groups of questions were designed to capture different aspects of these motivators.
But given the answers we can check whether they are indeed correlated to each other.

We used different questions to capture different aspects of the same motivator and grouped together questions related to the same motivator.
The grouping was content based.
Note that our grouping is subjective, and one could choose other related questions to the motivators.
We now examine how coherent are the groups that we choose, based on similarity on answers to questions of the same motivator.

based on their content, regardless of the actual answers to them.
In this section we examine how coherent the motivator is by measuring the Pearson correlation between the questions in the same motivator.
We first show what is the expected level of coherence.
We then compute the level of motivator coherence.
Last, we compare the coherence of the content-based factors to groups of questions deduced due to their high coherence.
}

As a reference of the level of correlation that we can expect, we focus on closely related question pairs.
For example, `I am skilled in software development' has a correlation of just 0.62 with `My code is of high quality'.
`I regularly reach a high level of productivity' and `I am a relatively productive programmer' have correlation of just 0.57.
Table \ref{tab:followup-factor-similarity} shows that the correlation of the same person answers to the motivation question in the original and follow-up survey is 0.52.
Note that a correlation of 0.5 is even higher than the correlation between LOC count and step functions on it \cite{amit2021follow}.
Therefore, coherence of about 0.6 is high.

\begin {table}[h!]\centering
\caption{ \label{tab:Follow-up-Coherence} Motivator Coherence }
\begin{tabular}{ | l| c| c| }
\hline
\textbf{Motivator} & \textbf{Coherence} & \textbf{Follow-up Coherence}\\
\hline
All Questions & 0.11 & 0.07\\ \hline
Community & 0.36 & 0.15\\ \hline
Enjoyment & 0.32 & 0.25\\ \hline
Hostility & 0.49 & 0.60\\ \hline
Ownership & 0.58 & 0.57\\ \hline
Recognition & 0.24 & 0.17\\ \hline
\end{tabular}
\end{table}

Table \ref{tab:Follow-up-Coherence} presents the coherence of the motivators.
`Coherence' is defined as the average Pearson correlation between all pairs of questions related to the same motivator.
The motivators `Challenge', `Ideology', `Importance', `Learning', `Payment', and `Self-use' do not appear in Table \ref{tab:Follow-up-Coherence} since they are based on a single question each, hence our method is not applicable to them.
`Follow-up Coherence' is the same metric as `Coherence' computed on the follow-up survey.
Note that this provides additional support, yet the support is not totally independent since the participants in the follow-up survey also participated in the original one.

The `All Questions' row represents all the questions together (basically related to motivation), and has rather low coherence.
The following motivators all have much higher coherence, indicating that they indeed reflect a meaningful grouping of questions related to specific concepts.
The coherence of `Hostility' and `Ownership' is relatively high in both surveys, and close to the highest coherence we can expect.
The coherence of `Community', `Enjoyment', and `Recognition' is moderate in both surveys.


\begin{figure}[!ht]
\hspace*{-20mm}\includegraphics[trim={0mm 0mm 0 0mm},scale=0.4]{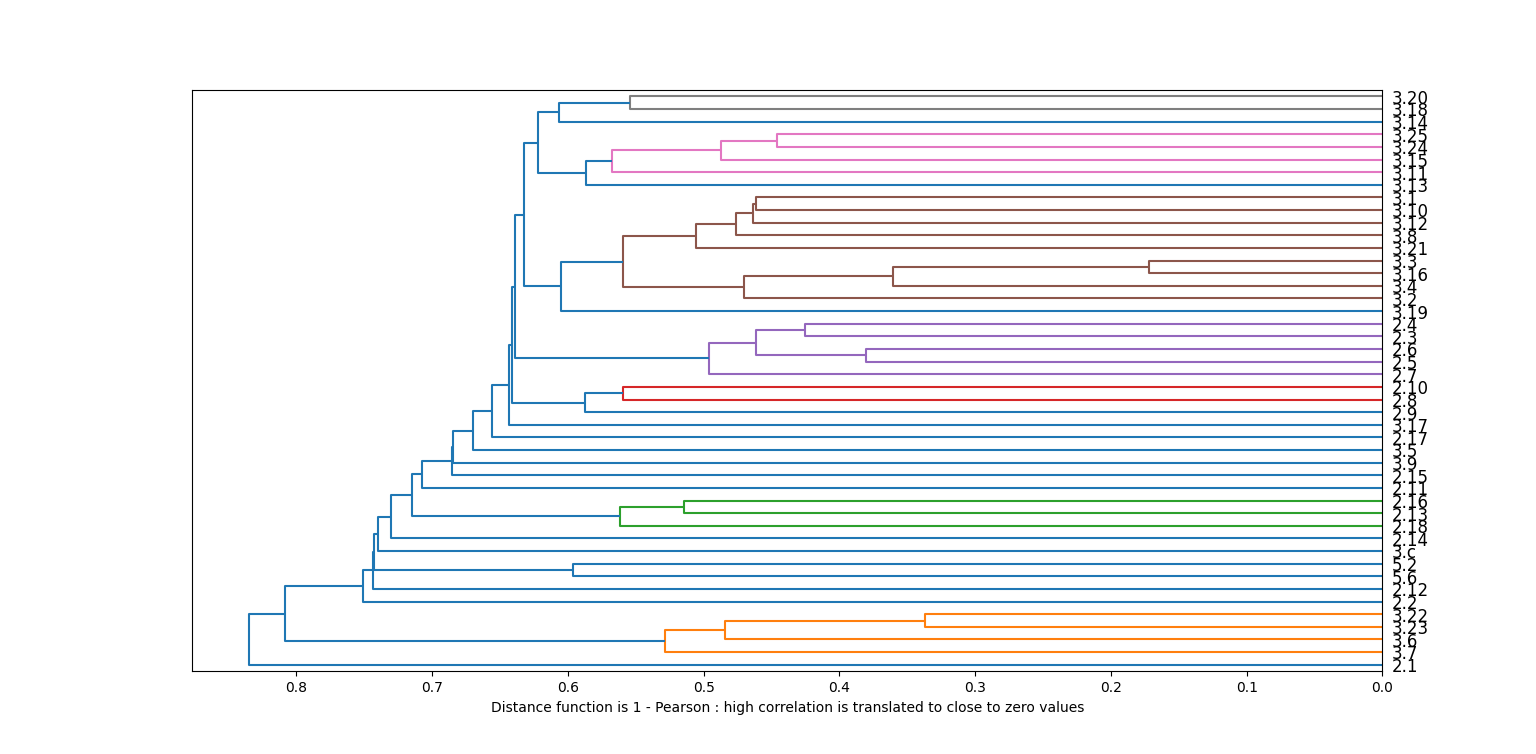}
\caption{\label{fig:questions_pearson_hc}
Dendrogram of questions based on their Pearson correlation. See questions text in \cite{replication} 
}
\end{figure}

Next, we compare the grouping of questions as we designed them with an automatic clustering based on their correlations.
This will provide additional evidence on whether our grouping was indeed meaningful.
The dendrogram in Figure \ref{fig:questions_pearson_hc} represents a hierarchical clustering \cite{murtagh2012algorithms} of questions based on the correlations between them.
The questions `I have significant influence on the repository' (3.3) and `I am a core member of the repository' (3.16) are the most correlated questions with Pearson of 0.83 (transformed to $1-0.83 =0.17$ to represent distance in the figure).
As we allow weaker correlations, more questions are clustered together, and when we allow Pearson correlation of only 0.3 most questions are already grouped into one big cluster, which is uninteresting.
Some of the clusters match our content-based motivators: 
The orange cluster matches hostility and the lower brown sub-cluster matches ownership.
The purple cluster shows a group that we did not consider: productivity and possible productivity improving elements such as good colleagues, physical conditions, and opportunities to use your abilities.
Other clusters may overlap our definitions, but also mix in unrelated questions.
For example, the pink cluster contains 3 questions about recognition and one about importance, which we feel is not really related.
Since our manually built factors are more coherent with respect to content, we use them and not the hierarchical clusters.

\subsection{Answers Stability Between Original and Follow-up Surveys}\label{sect:follow-up-stability}

The follow-up survey, conducted one year after the original survey, allowed us to compare the answers of the same person over time.
Table \ref{tab:followup-factor-similarity} shows stability of questions by motivator.

To compare the answers in the two surveys we first compute the Pearson correlation between them.
We also compute the differences between them, both the average absolute average difference (column `Avg.\ Abs.\ Diff') and the average relative difference (difference divided by the question value, column `Avg.\ Rel.\ Diff').
`Pred(25)' \cite{WEN201241} is the probability that the follow up answer is in the range of 25\% of the initial answer.

Note that the distributions of answers are far from uniform, and some answers are much more popular than others.
As a result, there is a high probability for getting the same answer even when the answers are independent.
`Pred(25) Lift' computes the lift, i.e.\ the extra probability above the expected Pred(25) from two independent answers from the answers distribution.

Pearson correlation, Pred(25), Pred(25) lift, and relative difference indicate stability for almost all motivators.
Hostility has a near zero lift and not a large positive one, indicating less stability than expected.
Note that the hostility distribution (e.g.\, Figure \ref{fig:q96.1_derogatory}) has a strong mode in the lowest value, making the independent distribution benchmark very high.
Note also that the lift is close to zero hence more likely to be influenced by noise.

\begin {table}[h!]\centering
\caption{ \label{tab:followup-factor-similarity} Similarity of Motivation Type Answers of Same Person in Two Dates }
\begin{tabular}{ | l| c| c| c| c| c| }
\hline
Motivator &         & Avg.\ Abs. & Avg.\ Rel. &          & Pred(25) \\
     & Pearson &    Diff    &  Diff      & Pred(25) &   Lift   \\
\hline
Learning & 0.68 & 0.91 & 0.04 & 0.81 & 0.22\\ \hline
Ownership & 0.66 & 1.02 & -0.01 & 0.83 & 0.42\\ \hline
Hostility & 0.63 & 1.10 & 0.38 & 0.38 & -0.02\\ \hline
Enjoyment & 0.60 & 1.06 & 0.00 & 0.84 & 0.29\\ \hline
Ideology & 0.57 & 1.61 & 0.02 & 0.74 & 0.99\\ \hline
Importance & 0.54 & 1.28 & 0.02 & 0.74 & 0.38\\ \hline
Motivation & 0.52 & 1.83 & -0.03 & 0.60 & 0.30\\ \hline
Challenge & 0.51 & 1.46 & 0.08 & 0.70 & 0.23\\ \hline
Community & 0.48 & 1.46 & 0.04 & 0.44 & 0.04\\ \hline
Recognition & 0.45 & 1.70 & 0.19 & 0.54 & 0.36\\ \hline
Self-use & 0.43 & 2.35 & 0.04 & 0.51 & 0.20\\ \hline
\end{tabular}
\end{table}

Payment is a binary feature hence its stability should be analyzed with different metrics.
The initial and follow-up payment agree in 85\% of the cases.
70\% of those that were paid in the initial survey were also paid a year later.
Only a single person out of the 27 that were not paid in the initial survey got payment in the follow-up.

Note that though the stability is moderate, it is higher than the one reported by Teghtsoonian and Teghtsoonian \cite{teghtsoonian1971repeatable}.
There, estimating the stability of the ``Stevens’s power law exponents'' which model human length perception, had correlation of about 0.4 after a few weeks and non-significant after a year.
Motivation is more abstract concept than length, more likely to raise ego defenses \cite{BassettJones2005Herzberg} and yet we showed it is more correlated.
Besides, over time the project, the people, and their motivations change \cite{gerosa2021shifting}, which might result in different answers.

\subsection{Face Validity of Answers}\label{sect:answer-validity}

To check the validity of the answers in our survey, we looked for mistakes, insincere answers, and biases.

The answer to the gender question was a free text field, in which the participant could write any answer.
Only 4 (0.8\%) of the answers had a typo (e.g., `mail', `boi').
Three of the answers were variants of `Attack Helicopter', a term ``used to disparage transgender people''\footnote{\url{https://en.wikipedia.org/wiki/I_Sexually_Identify_as_an_Attack_Helicopter}}.
Hence, these answers were probably not sincere.
In the country question, 1.2\% of the answers had a typo.

1.3\% of the developers said they had 15 years of experience with GitHub, established in 2008, which was impossible when the survey ended in 2021.
A single answer (0.2\%) of age of 100 years is probably insincere. 
Note that these error rates are much better than the 8.5\% who seemed to have given a wrong answer to a single simple question in \cite{herbold2022fine}, and the 10\% failure to identify negatively worded (reverse-coded) items discussed in \cite{podsakoff2003common}.

The job satisfaction questions were taken from a survey of 9,900 Australian clinical medical workers published in 2011 \cite{Hills2011Validation}.
Amusingly, software developers were on average less satisfied in all questions.
More importantly, questions about payment are irrelevant to volunteers and questions about community are irrelevant to people working alone.
We explicitly asked to skip these questions if they are irrelevant.
However, 57.7\% of the people that answered that they are not paid, answered the payment satisfaction question.
Therefore, it seems they answered regarding their salary from a different job, not related to the discussed project.
Some participants made comments in the open question that support this.

Some developers answered that they work on a public GitHub project.
For these projects, we checked the number of developers who committed code.
Of five developers who were found to work on single person projects, one answered most of the community-related questions, which are irrelevant to such projects.

\hide{
Both the original survey and the follow-up included the question `My personal motivation in this repository has increased since a year ago'.
80\% of the follow-up participants reported an increase in motivation.
However, their mean motivation answers actually slightly decreased by 0.052.
The point here is that most people reported an increase, contradicting their own answers.
}

There were questions in which the change in the follow-up survey is known in advance: age and experience should grow linearly with time.
The follow-up question was about a year after the first one.
In order to avoid rounding mistakes (e.g., a 20.5 years old participant might answer either 20 or 21), we consider answers as ``unreasonable'' only if the follow up answer was more than a year lower, or at least three years higher.
26\% of the answers about experience exhibited such unreasonable differences.
Two participants lost 5 years of experience each, somehow compensated by a participant that gained 11 years of experience in about a year.
16\% of the answers regarding GitHub experience were unreasonable.
However, for age, which has a higher presence in daily life, there were no unreasonable differences.

It seems that the biggest reliability problem comes from human failings \cite{podsakoff2003common}, bias due to ego defenses\cite{BassettJones2005Herzberg}, or the Dunning–Kruger effect (that people with lower capabilities tend to have higher self-esteem) \cite{Kruger99unskilledand}.
Only 5.6\% of the participants gave a low answer to `My code is of high quality', going up to 20.8\% when including neutral answers (6 on the 11-point scale).
The Pearson correlation with years of experience, a common method to estimate skill \cite{feigenspan2012measuring}, was a very low 0.06.
Moreover, first degree holders gave answers averaging 9.05, higher than all others.
People trained in computer science gave answers averaging 9.3, lower than the 10.5 average in math, yet a bit higher than arts (9.0), science (8.7), technology (8.6), and business (7.5).

Using the participants' GitHub profile, we can compare their actual activity to their self-perception.
People that answered that they write detailed commit messages (at least 9 - `somewhat agree'), had average commit message length of 89 characters, placing them in the 61 percentile of GitHub developers, not very far from the median.
Participants saying that they write high quality code have corrective commit probability (CCP) \cite{Amit2021CCP} of 0.36 (investing more than a third of their work in bug fixing), worse than 81\% of the GitHub developers.
We measured productivity by commits per working day \cite{47853, Oliveira2020TLProd}.
Developers that consider themselves to be productive contribute 3.14 commits per working day on average.
This is lower than the 3.34 commits of developers that do not consider themselves to be relatively productive.

It seems that there is also a bias leading to higher answers about the participant than about the community.
The average answer for questions about themselves is 9.1, 24\% more than the average answer to questions about the project.
A somewhat smaller difference of 4.5\% to 17.1\% was found when the questions were essentially paired (e.g, `My code is of high quality' and `The quality of the code in this repository is better than others').

We cannot accurately aggregate the probability of mistakes.
The probability of identifying an insincere answer is low, around 0.2\%.
Mistakes typically occur in few percent of the answers, yet in specific questions (the job satisfaction in our case) might be around 50\%.
Biases are the largest threat to validity, as demonstrated by the 79\% participants that consider their code to be of high quality.
Only 5.6\% of the participants gave a low answer to `My code is of high quality', and 20.8\% when including neutral.
The 5.6\% that think that their code is of low quality seem to be either more modest or more realistic.

\section{Threats to Validity}
\label{sect:threats}

Motivation, and motivators, are not well defined.
Therefore, it is hard to measure them or even evaluate how well a measurement method performs.
We cope with this threat using several methods like using questions from prior work, which were already considered to be useful.

The selection of motivators and questions has subjective aspects, and others could be chosen.
We based our selection on motivators with massive prior work in motivation in general, in software development, and open source.

Some questions have systematic problems.
The job satisfaction questions were answered by many participants on their day job.
In self-assessment questions developers have a very high perception of themselves, not aligned with their actual performance (Section \ref{sect:answer-validity}).
We therefore avoided using these answers for motivation analysis.

In order to further reduce the influence of individual questions, we grouped questions by motivators.
While we still do the analysis at the question level too (available in the supplementary materials), the aggregation reduces the weight of a specific answer and makes the concepts more robust.
However, answers to different questions on the same concepts are only moderately correlated (Section \ref{sect:motivating-factors}), so one can argue that our grouping is not correct.
Indeed, we grouped questions by subjective judgment of their contents, and in principle a different taxonomy could be used.
We compared our content based grouping to the one inferred from correlation (Section \ref{sect:motivating-factors}).
The match is only partial hence our grouping is supported yet there are also different justifiable groupings.

Surveys are answered by people.
Answers of the same person change over time (Section \ref{sect:follow-up-stability}) and therefore analysis based on the original survey might not agree with the same analysis on the follow-up survey.
Concerning the concept of hostility, different people provided widely different answers about the same project (Section \ref{sect:hostility}).
In this case this does not represent a data-quality problem, since the answers actually represent different experiences, and the difference itself is an important result.

We measured the relations between motivators and motivation in multiple ways: correlation, predictive performance, and co-change.
A similar result in all methods (e.g., community increases motivation by 20\%) would have been very reliable.
However, there are many quantitative and even several qualitative differences in the results.
For example, Table \ref{tab:high-answer-motivation-factor} shows that all motivators have positive precision lift in high motivation prediction, hence knowing of a positive motivator increases the probability of high motivation.
On the other hand, in the follow up analysis presented in Table \ref{tab:steps-motivation-factor}, three of the eleven motivators have negative precision lift, hence knowing of their increase from the original survey indicates higher probability of motivation reduction.
While negative lift is expected for hostility, the results for challenge and ideology disagree with the high motivation prediction.

Throughout this research we obtained many results.
While our number of participants is very high for a survey, we analyzed the answers in many ways.
In some scenarios (e.g., developers in the same project, developers answering the follow up), the numbers are quite small.
Statistical learning theory \cite{vapnik1971VC, vapnik2013nature} tells us that in such cases several of the empirical results will probably be different from the actual ones.
This is an inherited threat from the dataset size and analysis type, which should be resolved by replication studies obtaining more data and supporting the results in different analyses.

Similarity between the participants group and the desired population increases the probability of generalization.
We discussed in the demographics section (Section \ref{sect:stackoverflow}) that though our participants resemble the Stack Overflow survey participants (while being somewhat more professional), it is not clear what is the general developers group.



\section{Conclusions}
\label{sect:conclusions}

We conducted a large survey of software developers regarding motivation.
We grouped the questions by motivators and analyzed their relation to motivation.

Our supervised-learning-based analysis of motivators ranged from using each motivator as a classifier for motivation, through using motivator improvement as a classifier for motivation improvement, to constructing models that use the combined power of multiple motivators for improved prediction.
We confirmed that previously suggested motivators do indeed contribute to motivation.
At the same time, the influence of each individual motivator is limited, as also noted in prior work \cite{herzberg1986one, couger1988motivators}.

Apparently, the motivation of different developers, working in different contexts, may be influenced by different motivators.
No single motivator by itself is sufficient for inducing high motivation.
At the same time, none of the motivators is strictly necessary.
An analysis of the relations between them indicated that motivators tend to have low correlation.
This indicates that one should not look at motivators from the prism of which is the ``most important'' one;
a better description is that each one of them captures a different aspect of motivation \cite{mayer2007seventy}, and multiple aspects should be satisfied in order to have high motivation.


All motivators have coherence higher than the set of all questions together, but only hostility and ownership have rather high coherence.
In general, all motivators are at least moderately coherent and predictive, in all analyses.
However, out of the eleven motivators, ten motivators (excluding ownership) did not meet all three criteria of high coherence, stability, and predictive power.
This indicates that the bar that we set is high. 

It is also interesting to notice the relative position of payment.
Trying to predict high motivation based on a single motivator, payment has precision lift of 10\%, the lowest value of all positive motivators, and the only negative lift on the follow-up survey.
Since payment is the common way to promote motivation in businesses, it is important to note that other motivators might lead to a larger effect.

Hostility is a very coherent demotivator.
However, different people in the same project disagree on hostility, implying that it is not noticed by others.
Hence, not noticing hostility is not enough to assure lack of hostility in a project, and therefore actively looking for it might be needed.

\hide{Not related specifically to motivation, we quantified moderate reliability of answers.
People provide different answers to related questions, significantly change their answers over time, sometimes disagree with others in the same project, make errors, and overestimate themselves.
We therefore developed additional methods to overcome this.
We evaluated our content-based grouping of questions and compared it to automatic answer-based grouping.
We evaluated stability of answers over time and used the improvement in motivators' answers to predict improvement in general motivation.
We estimated the tendency of different types of errors and tendency of over-estimation.
By doing that, we made the reliability challenge more transparent and less likely to dispute the motivation results.
}

Participants who reported an improvement in the interest expressed in them had a large tendency for improvement in motivation.
Recognition, and specifically expressing interest, is free, applicable in all situations, and influential.
Considerate behavior and looking for practical benefits coincide here.
Be kind and give recognition, it is likely to pay off.

\section*{Data Availability}

All experimental materials (except for identifying data such as emails and GitHub profiles) is available at \cite{replication}.

\section*{Acknowledgments}

First, we would like to thank our participants.
Other than just answering the survey, they alerted on problems and suggested many ideas.
Many of them left their email to receive the research results which is heartwarming.
We could not conduct this research without you and you are our partners.
Thank you!

We thank the IRB committee for their feedback and approval of the study (09032020).
This research was supported by the ISRAEL SCIENCE FOUNDATION (grant no.\ 832/18).

We thank Davis Amit for the support, interest, and many wise ideas and questions.
We thank Avraham Kluger who introduced to us the importance of listening and helped in many ways, and to Dan Ariely who guided us into the domain of motivation.
We also thank Tali Kleiman, Itamar Gati, Zafrir Buchler, Doody Parizada, Eyal Zaidman, Yaniv Mama, Asaf Korem, and Nili Ben Ezra for the discussions, insights, and help.

\bibliographystyle{habbrv}
\bibliography{abbrv.bib,se.bib,bibtex.bib}

\appendix
\section{Survey Questions}
\label{sect:survey-questions}

To facilitate the review, the full questionnaire used in the survey is reproduced herewith.

\subsection{Survey introduction}

Dear participant,

We are a team of researchers interested in improving software development (see for example https://www.cse.huji.ac.il/\~{ }feit/papers/Refactor19PROMISE.pdf). 

If you contributed to a GitHub repository as a developer in the last 12 months, we ask for your help by answering questions about your contribution and motivation. 
Answering these questions is estimated to take 10--15 minutes of your time.

Based on the experience of respondents to this questionnaire in the past, you may gain new insights about your priorities in software development and areas of importance to you. 
Your answers, with the answers of others, will allow researchers in the future to investigate motivation, quality and productivity in software development and hopefully improve them.

We would appreciate a link to your GitHub profile in order to match your answers and GitHub activity (e.g., number of commits, years in the repository).
We are aware that the profile is a personal identifier and we will keep it private and use it for research purposes only.
The results of analysis of the profile data will be reported in aggregated form only.
Of course, in case that you are not interested, you can leave the field empty.

If you are willing to participate in this study, move to the next page. By moving to the next page you agree to participate in this study. The only inconvenience that this study may cause you is the need to concentrate on the questions for about 10-15 minutes. Yet, you may quit this survey at any time without answering all the questions, with no consequences for you. We will be grateful if you complete ALL the questions. 
No personally identifying information will be collected, except your GitHub profile if you choose to share it.

Thank you so much for your help.

Prof. Dror Feitelson, Prof. Avi Kluger,  Ph.D. candidate Idan Amit

 If you have any question you can contact Idan Amit at idan.amit@mail.huji.ac.il

\subsection{Questions regarding yourself}

The questions in this section are in Likert scale where 1 is `Strongly disagree' and 11 is `Strongly agree'.

\begin{enumerate}

    \item Productivity is more important to me than quality
    \item My motivation has more influence on my productivity, than my skill

    \item I regularly reach a high level of productivity  (based on \cite{47853})
	\item I am a relatively productive programmer

	\item I am skilled in software development (based on \cite{10.1007/978-3-319-33515-5_9})
	\item My code is of high quality 
	\item I am satisfied with my performance in software development \cite{10.1007/978-3-319-33515-5_9}
	\item I want my code to be beautiful 
	
	\item I enjoy software development very much

    \item It is important for me to program well (based on \cite{10.1007/978-3-319-33515-5_9})

	\item I write tests for my code

	\item I write detailed commit messages 
	\item I contribute to open source in order to have an online portfolio

	\item I try to write high quality code because others will see it
	\item I enjoy trying to solve complex problems \cite{Amabile94thework}

	\item I contribute to open source in order to become a better programmer

	\item I improved as a programmer since a year ago
	
	\item I contribute to open source due to ideology
	
\end{enumerate}

\subsection{Questions regarding activity in a repository}

Please choose \emph{one specific} GitHub repository that you work on.
Answer the following questions with respect to this repository. (These questions 

\begin{itemize}
	\item What is the link of the GitHub repository that you answer on? (Free ext)
	\item How many hours a week do you work on the repository (average)? (Free text)
	\item I’m being paid for my work in this repository (Yes/No)
\end{itemize}
\vspace{3mm}

The questions in this section are in Likert scale where 1 is `Strongly disagree' and 11 is `Strongly agree'.

\begin{enumerate}

	\item I regularly have a high level of motivation to contribute to the repository (based on \cite{47853})
	\item I have complete autonomy in contributing to the repository
        \item I have significant influence on the repository
	\item I feel responsible for the repository's success
	\item I’m interested in the repository for my own needs

	\item We have many heated arguments in the community. If you are the only developer in the project, please skip.
	\item I wish that certain developers in the project will leave. If you are the only developer in the project, please skip.

	\item My work on the repository is creative
	\item Working on this repository is challenging
	\item I derive satisfaction from working on this repository
	\item The repository is important
	\item When I look at what we accomplish, I feel a sense of pride.

	\item Belonging to the  community is motivating my work on the project. If you are the only developer in the project, please skip.
	\item The community is very professional. If you are the only developer in the project, please skip.
	\item I get recognition due to my contribution to the repository

	\item I am a core member of the repository

	\item I learn from my contributions

	\item The quality of the code in this repository is better than others 
	\item Code quality in the repository improved since a year ago 
	\item The project's community of developers is more motivated than that of other projects. If you are the only developer in the project, please skip.
    \item My personal motivation in this repository has increased since a year ago
	
	\item  In the past year, members of my project community put me down or were condescending to me. If you are the only developer in the project, please skip. (based on \cite{Cortina2001Incivility})
	\item  In the past year, members of my GitHub community made demeaning or derogatory remarks about me. If you are the only developer in the project, please skip.
 (based on \cite{Cortina2001Incivility})

    \item In the past year, members of my project community asked questions that show their understanding of my contributions. If you are the only developer in the project, please skip. (based on \cite{Kluger17FLS})
    \item In the past year, members of my project community expressed interest in my contributions. If you are the only developer in the project, please skip. (based on \cite{Kluger17FLS})

\end{enumerate}

\subsection{Job Satisfaction}

The following questions are from Job Satisfaction Scale questionnaire \cite{Hills2011Validation}.
We present the questionnaire as is in order to compare to previous results. 
In case that you find some questions irrelevant, please skip them.

The questions in this section are in Likert scale where 1 is `Extremely dissatisfied' and 7 is `Extremely satisfied', as in the original survey \cite{Hills2011Validation}.

The questions indicate level of satisfaction with the following:
\begin{enumerate}
    \item Freedom to choose your own method of working
    \item Amount of variety in your work
    \item Physical working conditions 
    \item Opportunities to use your abilities
    \item Your colleagues and fellow workers
    \item Recognition you get for good work
    \item Your hours of work 
    \item Your remuneration (payment) 
    \item Amount of responsibility you are given
    \item Taking everything into consideration, how do you feel about your work?
\end{enumerate}

\subsection{Demography}

\begin{enumerate}
	\item Country (Free text)
	\item Age (0-100 selection)
	\item Gender (Free text)
	\item I work as a professional programmer (Yes/No)
	\item Years of work experience (not including studies) (Free text)
	\item Years of contribution to GitHub(0-15 selection)
	\item Academic background (degree and graduation year) (Free text)
	\item Git profile link (Free text)
We would appreciate a link to your GitHub profile in order to match your answers and GitHub activity (e.g., number of commits, years in the repository).
We are aware that the profile is a personal identifier and we will keep it private and use it for research purposes only.
The results of analysis of the profile data will be reported in aggregated form only.
Of course, in case that you are not interested, you can leave the field empty.
\end{enumerate}

\subsection{Open questions}

\begin{enumerate}
	\item Do you have any comments on the questionnaire or research? Are you motivated due to a cause that we didn't consider? Do you have a method that increases your code quality? (Free text)
	\item Thank you for answering our survey. If you would like to be informed in the results of the research or to participate in the gift card lottery, please enter your email and we will send it to you once completed. The email will not be used for profile identification. (Free text)
\end{enumerate}

\end{document}